\documentstyle[preprint,eqsecnum,aps,epsf]{revtex}

\begin{document}
\title{SOFT PION EMISSION IN \\
SEMILEPTONIC $B$-MESON DECAYS}
\author{J. L. Goity}
\address{
Department of Physics, Hampton University, Hampton, VA 23668, USA \\
and \\
Continuous Electron Beam Accelerator Facility \\
12000 Jefferson Avenue, Newport News, VA 23606, USA.}
\author{
W. Roberts}
\address{
Department of Physics, Old Dominion University, Norfolk, VA 23529,
USA \\
and \\
Continuous Electron Beam Accelerator Facility \\
12000 Jefferson Avenue, Newport News, VA 23606, USA.}
\maketitle
\begin{abstract}
An analysis of semileptonic decays of $B$ mesons with the emission of
a single soft pion
is presented in the framework of the heavy-quark limit using an
effective Lagrangian  which implements chiral and heavy-quark
symmetries.
The analysis is performed at leading order of the chiral and inverse
heavy mass expansions. In addition to the ground state
heavy mesons some of their resonances are included. The estimates of
the various
effective coupling constants and  form factors needed in the analysis
are obtained using a chiral quark model.
As the main result, a clear indication is found that the $0^{+}$ and
$1^{+}$ resonances
substantially affect the decay mode with a
$D^{\ast}$ in the final state, and  a less  dramatic effect is also
noticed
in the $D$ mode. An analysis of the decay spectrum in the $D^{(
\ast)}-\pi$ squared invariant mass
is carried out, showing the main effects of including the resonances.
The obtained rates show
   promising  prospects for studies of soft pion emission in
semileptonic  $B$-meson  decays in a $B$-meson factory where, modulo
experimental cuts,
 about $10^5$ such decays in the $D$ meson mode and $10^4$ in the
$D^{\ast}$ mode
could be observed per year.
\end{abstract}
\pacs{  {\tt$\backslash$\string pacs\{13.25.Hw, 1.39.Fe, 14.40.Lb,
14.40.Nd \}} }
\newpage

\section{Introduction}

Semileptonic $B \rightarrow D^{(\ast)}$ decays with emission of a
single
pion may soon be established at CLEO or ARGUS, as well as at the
planned $B$-meson factory. These decays, denoted $B_{\ell 4}$ in the
rest of this article, complete the list of this category of decays
which includes $K_{\ell 4}$ and $D_{\ell 4}$. There are fundamental
differences
among these three decays, which make their study very interesting.
In particular, $K_{\ell 4}$ decays can be studied within chiral
perturbation theory ($\chi$PT)
over the whole final-state phase space as the resulting pions are
soft \cite{bijnens}.
 $D_{\ell 4}$ decays \cite{Dl4} are much harder to study because,
with the exception of a small fraction of phase space, the final
state
involves light mesons with relatively large energies. This makes the
use of an effective theory not viable.

 $B_{\ell 4}$ decays offer new theoretical possibilities.
As a whole, they are difficult to predict because the kinematic
domain of the
daughter pion ranges from the soft limit, to be properly defined
later,
to a high energy limit. However, it is possible to  restrict the
study  to
the soft-pion limit, where  one can make use of the  powerful
constraints resulting from heavy quark
 spin-flavor symmetry and chiral symmetry.

 In this work we study the soft-pion domain in
$B_{\ell 4}$ decays, for which the effective chiral Lagrangian
approach combined with the inverse heavy mass expansion \cite{CPTHM}
provides a consistent theoretical framework. In a strict sense, it
turns out that
the soft-pion domain represents about 80\% of the $B_{\ell 4}$ rate
in the $D$ mode and about 10\% in the $D^{\ast}$ mode.  Here the
large fraction in the D-mode
is mostly due to the inclusion of the cascade  decay
$B\rightarrow \ell \bar{\nu} D^{\ast}\rightarrow \ell \bar{\nu} D
\pi$.
Since, according to a rough estimate, the total branching ratio for
the $B_{\ell 4}$ decay is about 1\% (0.2 \%) in the $D$ ($D^{\ast}$)
mode, it seems
that experimental access to the soft-pion domain in the foreseable
future, such as at
a $B$ meson factory, is  clearly possible.

Perhaps the most compelling motivation for studying $B_{\ell 4}$
decays
is the overall current status of the semileptonic decays of $B$
mesons.
The measured inclusive semileptonic branching ratio is $11.1\pm 0.3$
\%,
while the sum of measured exclusive semileptonic branching fractions
is
significantly less than this. In particular, the elastic modes
$B\to D e\nu$ and $B\to D^* e\nu$ account for only about 60 \% of the
total semileptonic branching fraction \cite{bellantoni} ($b\to u$
modes
are expected to be suppressed by $|V_{bu}|^2/|V_{bc}|^2\approx 0.01$).
Understanding the so-called inelastic modes, such as those we discuss
here, is therefore crucial to resolving the apparent discrepancies
among the measurements.

Another  interesting aspect of $B_{\ell 4}$-decays is that they may
give an indication of resonance effects (especially  $D$ meson
resonances).
At present, the only well established resonances are the two $P$-wave
objects $D_1$ and $D_2$, with $J^P=1^+,\,2^+$, respectively.
States that may contribute significantly to $B_{\ell 4}$ decays, and
hence which may be observed in such decays, include the remaining
lowest-lying $P$-wave states (with $J^P=0^+,\,1^+$), two of the
lowest lying $D$-wave states ($J^P=1^-,\,2^-$), and the first
radially
excited $S$-wave states
($J^P=0^-,\,1^-$). It turns out that the well established $D_1$ and
$D_2$ states \cite{resonances} do not play a role in the present
analysis, while only the $S$-wave radially excited $D$ mesons offer
any opportunity for direct discovery using the $B_{\ell 4}$ decays,
as they are the only ones with a small enough width to show up as a
resonance feature in the decay spectrum.

{}From the theoretical standpoint, the soft-pion limit is  particularly
interesting, as chiral symmetry and spin-flavor symmetry  determine
 to a large extent the different decay amplitudes, in terms of a few
low-energy constants and universal form factors. These form factors
 are associated with the matrix elements of the charged $b\rightarrow
c$ electroweak current between the relevant heavy meson states,
 and the low-energy  constants determine the amplitudes of the strong
interaction transitions mediated by the soft pion.

Another area of interest in $B_{\ell 4}$ decays is their contribution
to $\rho$, the slope of the Isgur-Wise function which, in turn,
 impacts on the extraction of $|V_{cb}|$ from data. Bjorken, Dunietz
and Taron \cite{Taron} have obtained a sum rule that relates the
slope of
 the Isgur-Wise function for the elastic decays $B\to D\ell\nu$ to
the form factors that describe the inelastic semileptonic decays.
 The decays that we consider here provide the leading resonant and
non-resonant contribution to this quantity.

These decays have been analyzed in the framework we explore by a
number of authors. $B\to D\pi e\nu$ has been treated by Lee,
Lu and Wise \cite{wiselee}, and by Cheng {\it et al.}
\cite{chengetal}, while Lee \cite{lee} and Cheng {\it et al.}
 \cite{chengetal} have examined $B\to D^*\pi e\nu$. In these
analyses, only the  ground state mesons,
 the $D$, $D^*$, $B$ and $B^*$ were included, which amounts to
keeping only those contributions wich are leading
at the zero recoil point  $v\cdot v^\prime=1$, $v$ and $v^{\prime}$
being the four velocities of the $B$ and $D$ mesons,
respectively.  In addition, we note that only Cheng {\it et al.} went
on to estimate branching fractions and evaluate the
differential decay spectra.

In this work we include resonances which contribute at leading order
in the chiral expansion, and neglect those which correspond
 to radially excited states (except the  two resonances $0^{-}$ and
$1^{-}$ which we have to keep as shown later).
 Since resonance contributions vanish at zero recoil in the infinite
heavy quark mass limit, their effects   can only
 be observed away from that point, and as we will demonstrate, they
alter substantially the decay rates.

The predictions that arise in our analysis rely heavily on our
ability to obtain good estimates of the aforementioned
 low-energy constants and universal form factors. In this work we use
the chiral quark model \cite{chiralquarkmodel}
convoluted with wave functions obtained in a simple model of heavy
mesons  to give such estimates. It is our hope that
 this procedure leads to reasonable results.

The experimental status of $B_{\ell 4}$ decays is hazy. ARGUS has
analyzed the process
 $\bar{B}\to D^{**}\ell^-\bar\nu$ as background to the semileptonic
decays $B\to D\ell\nu$ and $B\to D^*\ell\nu$.
 The resonances included in their analysis were the four $P$-wave
states alluded to above,
as well as the two radially-excited $S$-wave states. They report $63
\pm 15\pm 6$ possible candidates \cite{argus1}.
 This result is based on studying the invariant mass distribution of
the $D^*\pi$ combinations that result from the decay of the $D^{**}$.
The corresponding branching ratios are $BR(\bar{B^0}\to D^{**}\ell^-
\bar\nu)=(2.7\pm 0.5\pm 0.5)\%$, when their results are fitted to the
model
 of Isgur, Scora, Grinstein and Wise (ISGW) \cite{isgw}, and $BR(
\bar{B^0}\to D^{**}\ell^-\bar\nu)=(2.3\pm 0.6\pm 0.4)\%$ when fitted
to the
model of Ball, Hussain, K\"orner and Thompson (BHKT) \cite{bhkt}.
This result implies that the resonant contribution to $B_{\ell 4}$
decays is significant. In the case of the decay $B\to D\pi\ell\nu$,
the $D^*$ provides the dominant contribution, largely because of its
proximity to the $D\pi$ threshold. As far as we know, no other
experimental group has published numbers for semileptonic decay rates
of $B$ mesons to excited $D$ mesons.

This article is organized as follows. In section II we review the
effective theory resulting from
spin-flavor and chiral symmetries. Section III presents the
calculation of the effective coupling constants and form factors
appearing in the effective theory, while we present the analysis of
the decay amplitudes and differential widths in section IV. Section V
is devoted to the results and discussions. A number of calculational
details are relegated to three appendices.

\section{Effective Theory }

In this section we briefly review the effective theory  which
incorporates
simultaneously spin-flavor and chiral symmetry \cite{CPTHM} and
describes the interactions between soft pions and heavy mesons.
Within the framework of the heavy quark effective theory (HQET)
\cite{Isgwise} , a hadron of total spin $J$ consists of a light
component (the brown muck) with spin $j$, and
the spin-1/2 heavy quark, with $J=j\pm 1/2$. For a given $j$, there
are therefore two mesons, and these are degenerate members of a
spin-flavor multiplet, at
leading order in HQET. In the rest of this article, we denote the
multiplets by the $J^P$ of the two states. For example, for
$j^P=1/2^-$, we have the multiplet
$(0^-,1^-)$.

For reasons that we will outline later, the only multiplets of
interest in this work are $(0^{-},1^{-})$, $(0^{+},1^{+})$,
$(1^{-},2^{-})$, and $(0^{-},1^{-})^{\prime}$.
Here, $(0^{-},1^{-})^{\prime}$ denotes the first radially excited
version of the ground state $(0^{-},1^{-})$ multiplet.
In order to formulate the effective theory, it is very convenient to
introduce
superfields associated with each multiplet \cite{bjorken}. These
superfields
provide a natural way of realizing the spin-flavor symmetry. At
leading order in the inverse heavy mass expansion one associates one
such
superfield with each possible four velocity $v_{\mu}$. This is
because in the large-mass
limit of the heavy quark, a velocity superselection rule sets in
\cite{georgi}.

The superfield assigned to the ground-state heavy meson multiplet
$(0^{-},1^{-})$
 with velocity $v_{\mu}$ is
\begin{equation}
{\cal H}_{-}=\frac{1+\not{\! v}}{2}\;\left(-\,\gamma_{5}\,P+\gamma^{
\mu}
V_{\mu}^{\ast}\right),
\end{equation}
where $P$ and $V_{\mu}^{\ast}$ ($v^{\mu}V_{\mu}^{\ast}=0$)
are the fields associated with the pseudoscalar
 and vector partners, respectively. These fields contain annihilation
operators only, and are obtained from the relativistic fields as
\begin{eqnarray} \label{groundstate}
P(x)&=& \sqrt{M}\;e^{-i \,M\, v     \cdot x}\;\Phi^{(+)}(x),\nonumber
\\
V^{\ast}_{\mu}(x)&=& \sqrt{M}\;e^{-i \,M\,v     \cdot x}\;
\Phi^{(+)}_{\mu}(x),
\end{eqnarray}
where the label (+) refers to the positive frequency modes of the
relativistic field, and $M$ is the meson mass.

The spin-symmetry transformation law is
\begin{eqnarray} \label{spinsymmetry}
{\cal H}_{-}&\rightarrow &\exp{\left(-i\,\overrightarrow{\epsilon}
  \!\cdot\!\overrightarrow{S}_{v}\right)}{\cal H}_{-},\nonumber \\
S_{v}^j&=&i\,\epsilon^{jkl}\,[{\not \! e}_{k},\not{\! e}_{l}]\,
\frac{(1+\not{\! v})}{2}.
\end{eqnarray}
where $e_{k}^{\mu}$, $k=1,2,3$, are space-like vectors orthogonal to
the four-velocity.
$\bar{\cal H}_{-}=\gamma_{0}{\cal H}_{-}^{\dagger}\gamma_{0}$
 transforms contravariantly to
${\cal H}_{-}$.

In a similar manner, it is straightforward to define the superfields
associated
with the excited states \cite{falk}. In our case, the states of
interest
are the
  $(0^{+},1^{+})$,  $(1^{-},2^{-})$ and  $(0^{-},1^{-})^{\prime}$
multiplets, which are
described by the superfields
\begin{eqnarray} \label{superfields}
{\cal H}_{+}&=&\frac{1+\not{\! v}}{2}\;\left(- H_{0^+}
+\gamma_{\mu}\gamma_{5} H_{1^+}\mu\right),\nonumber \\
{\cal H}_{-}^{\mu}&=&\frac{1+\not{\! v}}{2}\;\left(-\sqrt{\frac{3}{2}}
H_{1^{-}}^{\prime\,\nu}\left[g^{\mu}_{\nu}-\frac{1}{3}
\gamma_{\nu}(\gamma^{\mu}+v^{\mu})\right]+\gamma_{5}\gamma_{\nu}
\,H_{2^{-}}^{\mu\nu}
\right),\nonumber \\
{\cal H}_{-}^{\prime}&=&\frac{1+\not{\! v}}{2}\;\left(-\gamma_{5}
H_{0^-}^{\prime}
+\gamma_{\mu}H_{1^-}^{\prime\mu}\right),
\end{eqnarray}
respectively.
 All the tensors are transverse to the four-velocity, traceless and
symmetric.
 The transformations of  these superfields under spin symmetry
operations
are implemented in exactly the same manner as in the case of ${\cal
H}_{-}$.

The chiral transformation law of the superfields is easily determined
by following the well known Coleman-Wess-Zumino procedure to
implement non-linear realizations
of non-Abelian symmetries. All multiplets
are isodoublets (we do not include the $s$-quark in our analysis),
so that the transformation law under an arbitrary chiral rotation
belonging to $SU(2)_{L}\bigotimes SU(2)_{R}$ is
\begin{equation}\label{chiraltransformation}
{\cal H}\rightarrow   h(L,R,u){\cal H},
\end{equation}
where ${\cal H}$ is any spin symmetry multiplet, and $h(L,R,u)$
is an $SU(2)$ matrix which results from solving the system of
equations
\begin{eqnarray} \label{chiralsystem}
L\,u&=&u^{\prime} h(L,R,u),\nonumber \\
R\,u^{\dagger}&=&u^{\prime \dagger} h (L,R,u).
\end{eqnarray}
Here, $L$ ($R$) is an $SU_{L}(2)$ ($SU_{R}(2)$) transformation and $u$
is given in terms of the Goldstone modes (pions) as
\begin{eqnarray}\label{goldstone}
u(x)&=&\exp{\left(-\frac{i}{2\,F_{0}} \Pi(x)\right)},\nonumber \\
\Pi(x)&\equiv&\overrightarrow{\pi}     \!\cdot\!\overrightarrow{
\tau},~~~~F_{0}=93 {\rm MeV}.
\end{eqnarray}
The transformation (\ref{chiraltransformation}) is like a gauge
transformation, the $x$-dependence entering
via the Goldstone boson field. In order to build an effective
Lagrangian
which is chirally invariant, a covariant derivative is thus required,
and is
\begin{eqnarray}\label{covariantderivative}
\nabla_{\mu}&=&\partial_{\mu}-i\,\Gamma_{\mu},\nonumber \\
\Gamma_{\mu}&=&\Gamma_{\mu}^{\dagger}=\frac{i}{2}\,(u\partial_{\mu}
u^{\dagger}+
u^{\dagger}\partial_{\mu} u)=\frac{i}{8F_{0}}\,\left[ \Pi,\;
\partial_{\mu}\Pi\right]+\dots.
\end{eqnarray}

Another  fundamental element in the construction of the effective
Lagrangian is the
pseudovector,
\begin{equation}\label{pseudovector}
\omega_{\mu}=\frac{i}{2}\,(u\partial_{\mu} u^{\dagger}-
u^{\dagger}\partial_{\mu} u)=\frac{1}{2F_{0}}\partial_{\mu}\Pi+\dots,
\end{equation}
which transforms homogeneously under chiral transformations.

Since the rest mass of the different heavy mesons is removed
according to equations analogous to
eqn. (\ref{groundstate}), $\nabla_{\mu}$ acting on the respective
superfields is proportional to the residual momentum
carried by the superfield, which in the present analysis is  of the
order of
the pion momentum. This implies that $\nabla_{\mu}$
counts as a quantity of ${\cal O}(p)$ in the chiral expansion.
Obviously,
$\omega_{\mu}$ is of the same order.

Throughout this work we consider only the leading terms in the
inverse heavy mass expansion and in the chiral expansion, and neglect
the effects of chiral symmetry
breaking  due to the light quark masses (they enter only via the pion
mass
when the final-state phase space is considered). As we point out
later, we  have to include the effects of hyperfine splitting in the
ground-state
mesons, even though this splitting represents a departure from the
spin-flavor symmetry. These modifications are only kinematic.

To leading order in $\chi$PT the amplitudes for the $B_{\ell 4}$
decays are proportional
to a single power of the pion momentum. This places a   restriction
on the angular momentum
quantum numbers of the excited states that can be considered in this
analysis. The
excited states that contribute to the $B_{\ell 4}$ decay amplitudes
at this order can be identified
by examining their decays, via soft-pion-emission, to the ground
state heavy mesons.
Consider an excited state with spin $J=j\pm 1/2$, where $j$ is the
spin of the light component of the meson
(the brown muck). Let us define the integer $k\equiv j-1/2$. The
soft-pion decay amplitude of such a state to the
ground state supermultiplet is of ${\cal O}\left(p^k\right)$ if the
parity of the excited meson
is $(-1)^k$, or of ${\cal O}\left(p^{k+1}\right)$ if the parity is
$(-1)^{k+1}$. In the case where
$k=0$ and the parity of the resonance is positive, as is the case of
the
$(0^{+}, 1^{+})$ multiplet, the amplitude is proportional to  $p\cdot
v$ and is of order $p$ as well.
Using these `rules', one finds that only the states with the quantum
numbers mentioned above contribute at leading order in $\chi$PT.

The effective theory is
given by an effective Lagrangian which explicitly displays spin-flavor
and chiral symmetry. The ${\cal O}(p)$  strong interaction  part is
\begin{eqnarray} \label{stronginteraction}
{\cal L}_{\chi}&=&{\cal L}_{\chi}^{GB}+{\cal L}_{\chi}^{{\cal H}_{-}}+
{\cal L}_{\chi}^{{\cal H}_{+}}+{\cal L}_{\chi}^{{\cal H}_{-}^{\mu}}+{
\cal L}_{\chi}^{{\cal H}_{-}^{\prime}}
+{\cal L}_{\chi}^{{\rm int}},\nonumber\\
{\cal L}_{\chi}^{GB}&=&-\frac{F_{0}^{2}}{4}\;{\rm Tr}\left(\partial_{
\mu} U\partial^{\mu}
U^{\dagger}\right)-\frac{1}{2} B \,F_{0}^{2}\; {\rm Tr}\left({\cal
M}(U+U^{\dagger})\right)
+{\cal O}(p^{4}),\nonumber \\
{\cal L}_{\chi}^{{\cal H}_{-}}&=&-\frac{i}{2}\, v_{\mu} {\rm Tr_{D}}
\left( \bar{\cal H}_{-}
\stackrel{\leftrightarrow}{\nabla}^{\mu}{\cal H}_{-}\right)+
g \,{\rm Tr_{D}}\left( \bar{\cal H}_{-}\,\omega^{\mu}{\cal H}_{-}
\gamma_{\mu}\gamma_{5}\right)+
{\cal O}(p^{2}),\nonumber\\
{\cal L}_{\chi}^{{\cal H}_{-}}&=&-\frac{i}{2}\, v_{\mu} {\rm Tr_{D}}
\left( \bar{\cal H}_{+}
\stackrel{\leftrightarrow}{\nabla}^{\mu}{\cal H}_{+}\right)+\dots
\nonumber \\
{\cal L}_{\chi}^{{\cal H}_{-}^{\mu}}&=&-i v_{\mu} {\rm Tr_{D}}
\left( \bar{\cal H}^{\nu}_{-}
\stackrel{\leftrightarrow}{\nabla}^{\mu}{\cal H}_{-\nu }\right)+\dots
\nonumber \\
{\cal L}_{\chi}^{{\cal H}_{-}^{\prime}}&=&-\frac{i}{2} v_{\mu} {\rm
Tr_{D}}\left( \bar{\cal H}^{\prime}_{-}
\stackrel{\leftrightarrow}{\nabla}^{\mu}{\cal H}^{\prime}_{-}\right)+
\dots .
\end{eqnarray}
$U(x)=u^{2}(x)$ and ${\rm Tr_{D}}$
is the trace over Dirac indices. The interaction terms involving
$\omega_{\mu}$ have not been displayed in the Lagrangians of the
resonant states,
as they are not needed in this work.
The only interaction terms in ${\cal L}_{\chi}^{{\rm int}}$ we need
are those
which give transitions
between the resonances and the ground state mesons via a single soft
pion
emission. These are characterized by three low energy constants
$\alpha_{1}$, $\alpha_{2}$ and $\alpha_{3}$, and are
\begin{eqnarray} \label{interactionterms}
{\cal L}_{\chi}^{{\rm int}}&=&\alpha_{1}\;{\rm Tr_{D}}
\left( \bar{\cal H}_{+}\,\omega^{\mu}{\cal H}_{-}\gamma_{\mu}
\gamma_{5}\right)
+\alpha_{2}\;{\rm Tr_{D}}
\left( \bar{\cal H}_{-^{\mu}}\,\omega^{\mu}{\cal H}_{-} \gamma_{5}
\right)\nonumber\\
&+&\alpha_{3}\;{\rm Tr_{D}}
\left( \bar{\cal H}_{-}^{\prime}\,\omega^{\mu}{\cal H}_{-}\gamma_{
\mu}\gamma_{5}\right).
\end{eqnarray}
The vertices resulting from ${\cal L}_{\chi}$   needed in this work
are displayed in Appendix A.

The range of validity of the soft-pion limit is estimated from the
invariant product of the pion momentum and the four velocity.
As long as this product is smaller than some scale $\Lambda_{\chi}$,
which is of the order of 0.5 GeV, the  application of the soft-pion
 limit should be appropriate. In this limit, this product has to
remain smaller than the mass splittings between the neglected
resonances and the ground state mesons, thus giving
an estimate of the value of $\Lambda_{\chi}$. Since two velocities
appear, namely the velocities of
the parent $B$ meson and the daughter $D$ meson, the  soft-pion limit
requires that
the invariant products of the pion momentum with both velocities
must be smaller than  $\Lambda_{\chi}$

Another set of essential ingredients are the matrix elements of the
electroweak charged current $\bar{c}\gamma_{\mu}(1-\gamma_{5})b$.
Since this current
is an isosinglet,  it does not have direct couplings to pions at
leading chiral
order, and its matrix elements are easy to parametrize in the
effective theory.
Denoting the superfields corresponding to $D$ and $B$ mesons and
their resonances
respectively by ${\cal D}$ and ${\cal B}$, the matrix elements of the
charged current are obtained from the effective current operators
\begin{eqnarray}\label{currents}
({\rm a})~~~~~~~J_{\mu}((0^{-},1^{-})\rightarrow(0^{-},1^{-}))&=&
\xi(\nu)\, {\rm Tr_{D}}\left( \bar{\cal D}_{-}(v^{\prime})\gamma_{
\mu}(1-\gamma_{5})
{\cal B}_{-}(v)\right),\nonumber\\
({\rm b})~~~~~~~J_{\mu}((0^{+},1^{+})\rightarrow(0^{-},1^{-}))&=&
 \rho_{1}(\nu)\,\left[ {\rm Tr_{D}}\left( \bar{\cal D}_{+}
(v^{\prime})\gamma_{\mu}(1-\gamma_{5}){\cal B}_{-}(v)\right)\right.
\nonumber\\
&+&\left.{\rm Tr_{D}}\left( \bar{\cal D}_{-}
(v^{\prime})\gamma_{\mu}(1-\gamma_{5}){\cal B}_{+}(v)\right)\right],
\nonumber\\
({\rm c})~~~~~~~J_{\mu}((1^{-},2^{-})\rightarrow(0^{-},1^{-}))&=&
 \rho_{2}(\nu)\,\left[v_{\rho} {\rm Tr_{D}}\left( \bar{\cal D}_{+}^{
\rho}
(v^{\prime})\gamma_{\mu}(1-\gamma_{5}){\cal B}_{-}(v)\right)\right.
\nonumber\\
&+&\left. v^{\prime}_{\rho}{\rm Tr_{D}}\left( \bar{\cal D}_{-}
(v^{\prime})\gamma_{\mu}(1-\gamma_{5}){\cal B}_{+}^{\rho}(v)\right)
\right],
\nonumber\\
({\rm d})~~~~~~~J_{\mu}((0^{-},1^{-})^{\prime}
\rightarrow(0^{-},1^{-}))&=&
\xi^{(1)}(\nu)\, {\rm Tr_{D}}\left( \bar{\cal D}_{-}^{\prime}(v^{
\prime})\gamma_{\mu}(1-\gamma_{5})
{\cal B}_{-}(v) \right.\nonumber\\
&+&     \left.  \bar{\cal D}_{-}(v^{\prime})\gamma_{\mu}(1-\gamma_{5})
{\cal B}_{-}^{\prime}(v) \right).
\end{eqnarray}
Here, $v$ ($v^{\prime}$)  is the four-velocity of the $B$ ($D$)
meson, and
$\nu\equiv v \cdot  v^{\prime}$. $\xi(\nu)$ is the Isgur-Wise form
factor, which
is normalized to be unity at zero recoil ($\nu=1$) if one ignores
QCD corrections and higher orders in the  inverse heavy mass
expansion. The
other form factors, namely $\xi^{(1)}(\nu)$, $ \rho_{1}(\nu)$ and $
\rho_{2}(\nu)$,
 which  determine the transition between the resonances
and ground state mesons via the charged current, are not constrained
by symmetry at zero recoil. The currents however do vanish at zero
recoil
due to kinematic factors. In fact, if a resonance  is characterized by
a given value of the previously defined parameter $k$, the current
of interest is suppressed by a kinematic factor
of the form $(v-v^{\prime})_{\mu_{1}}\,(v-v^{\prime})_{\mu_{2}}...
\,(v-v^{\prime})_{\mu_{k}}$.
Note that no suppression of this form appears for the current of eqn.
(\ref{currents}(d)). However, heavy
quark symmetry and orthogonality together imply
that $\xi^{(1)}(\nu)$ has to vanish at zero recoil, at least as $(
\nu-1)$.
The expressions for the currents of eqn. (\ref{currents}) are given
in Appendix B.

We note that by choosing to work to leading order in $p$, we have
also placed an implicit restriction on the
powers of $\nu-1$ that appear. Since the largest value of $k$ to be
considered is unity,
 we find that the amplitudes for the $B_{\ell 4}$ decays are
proportional
to at most a single power of $v-v^\prime$, and  the differential
decay width
will contain terms with   at most two powers of
$\nu-1$.

We conclude this section by making a final comment on the states we
include in our analysis.
As outlined above, working to order $p$ has severely restricted the
states we can include, at least as far as their angular
momentum quantum numbers go. However, there is no restriction on
their `radial' quantum numbers.
Our self-imposed restriction of excluding any radially excited states
(with the exception of the radially excited
$(0^-,1^-)^\prime$ multiplet) is motivated by two related factors.
These radially excited states are expected to be quite a bit more
massive than their non-radially-excited counterparts.
Thus, we expect little contribution from such states, provided we do
not venture too far from the
non-recoil point. Furthermore, the propagators of such states are
expected to lead to a further
suppression of any contribution, as in the strict soft pion limit
these states will be far off their mass shell.

\section{LOW ENERGY CONSTANTS AND FORM FACTORS}

The soft-pion limit of the $B_{\ell 4}$ decays is determined
in terms of four low energy constants: $g$, $\alpha_{1}$, $
\alpha_{2}$ and $\alpha_{3}$;
four universal form factors: $\xi(\nu)$, $\xi^{(1)}(\nu)$, $
\rho_{1}(\nu)$ and $ \rho_{2}(\nu)$;
and mass differences between the resonances and the ground state
mesons: $\delta m_{1}\equiv M_{0^{+}}-M_{0^{-}}$,
$\delta m_{2}\equiv M_{2^{-}}-M_{0^{-}}$, and $ \delta m_{3} \equiv
M_{0^{\prime -}}-M_{0^{-}}$, and
the total decay widths of the resonances. In writing this form, we
are treating the states in each excited
multiplet as degenerate with each other. However, in dealing with the
contribution from the ground-state doublet $(0^-,1^-)$, it is
imperative that
we include the mass difference $\delta m_{0}\equiv
M_{1^{-}}-M_{0^{-}}$, as this plays a profound role on the outcome of
our analysis.
 We begin by formulating a simple model of the heavy mesons, and
using the wave functions obtained from
this model to calculate the quantities we need.

To estimate the masses and obtain wave functions, we use a version of
the Godfrey-Isgur model \cite{gi}. In this version, one of the quark
masses is set to infinity. In addition, we do not expand the wave
function of a state in a harmonic-oscillator basis, but instead choose
it to be that of a single harmonic-oscillator state with the
appropriate quantum numbers. The oscillator parameter of each wave
function is obtained in one of two ways. In the first method, we
perform a variational calculation, and the values obtained in this way
are listed in table \ref{qmparameters}. Also listed in this table are
the mass differences between the excited states and the ground states.
We discuss the second method of obtaining the value of the gaussian
parameter below.

The low energy constants appropriate for soft-pion emission are
estimated in a chiral quark model. In this model, pions couple only to
the light
constituent  quarks
of a heavy hadron, via the Lagrangian
\begin{eqnarray}\label{chiralquark1}
{\cal L}&=&i\bar q\gamma^\mu\nabla_\mu q-\tilde m_q\bar q
q+g_A^q(0)\bar q\gamma_\mu\gamma_5\omega^\mu q,\nonumber\\
\nabla_\mu q&=&\partial_\mu q +i\Gamma_\mu q,
\end{eqnarray}
where $\Gamma_{\mu}$ and $\omega_{\mu}$ were defined previously,
$\tilde m_q$ is the constituent quark mass,
and the constituent quark field $q$ transforms under chiral rotations
as $q\to h(L,R,u)q$. The axial coupling of the quark, $g_A^q(0)$, is
assumed to be unity. Arguments  favoring this value for the axial
coupling constant have been given in \cite{weinberg}.
A non-relativistic expansion of the interaction term of this
Lagrangian
is performed, and the resulting non-relativistic interaction term is
convoluted with wave functions obtained from the model described
previously. The low energy constants obtained in this way are also
listed in
table \ref{qmparameters}. More details of this model will be presented
elsewhere.

One of the results of this analysis is that some of the low energy
constants vanish in the limit when the pion energy in the vertex is
taken to
zero. For the cases where this chiral suppression occurs, we define
the
low energy constants as corresponding to the energy of the pion
in the decay of an on-shell heavy resonance to an on-shell heavy
ground-state.
We expect that this procedure will furnish only rough estimates  of
these couplings.

We also use the chiral quark model to estimate the total   and partial
widths of the excited states relevant to our analysis. Phase space
limits all decays
to pions or $\eta$'s, both of which can be described in terms of
chiral
dynamics. It turns out that decays to $\eta$'s play only a small role,
and then only for the $(1^-,2^-)$ multiplet
($\Gamma_{(1^-,2^-)\to(0^-,1^-)\eta}\approx$ 4 MeV).

We can compare the results we obtain for the partial and total widths
of these states with the calculation of Godfrey and Kokoski \cite{gk},
for instance.
Unfortunately, we have only a single pair of states in common with
that
calculation, namely the $(0^+,1^+)$ multiplet. The large total widths
we obtain
for these states are consistent with the widths obtained in
\cite{gk}.

The low energy constants $g$, $\alpha_{1}$, $\alpha_{2}$ and
$\alpha_{3}$ are related to the partial  widths
for the resonance decays into  ground state mesons via single pion
emission by
\begin{eqnarray}\label{widthsandconstants}
\tilde{\Gamma}_{1^{-}}&=&\frac{g^{2}}{8\pi
F_{0}^{2}}\,p_\pi^3,\nonumber\\
\tilde{\Gamma}_{0^+}=\tilde{\Gamma}_{1^+}&=&\frac{3\alpha_1^2}{8\pi
F_0^2}\frac{M_D}{M_D+\delta m_1}E_\pi^2p_\pi,\nonumber\\
\tilde{\Gamma}_{1^-}=\tilde{\Gamma}_{2^-}&=&\frac{\alpha_2^2}{8\pi
F_0^2}\frac{M_D}{M_D+\delta m_2}p_\pi^3,\nonumber\\
\tilde{\Gamma}_{0^{-\prime}}=\tilde{\Gamma}_{1^{-\prime}}&=&\frac{
\alpha
_3^2}{8\pi F_0^2}\frac{M_D}{M_D+\delta m_3}p_\pi^3.
\end{eqnarray}
It is interesting to note that the value $g=0.5$ obtained here is a
direct result
of the assumption that $g_{A}^{q}(0)=1$ and is independent of the wave
function used.

The total widths of these states are similar to the partial widths
obtained in this fashion, with one exception. The total width of the
 $(1^-,2^-)$ resonances is dominated by their decay into the
$(1^+,2^+)$ resonances, with a resulting total width of 405 MeV.

The form factors $\xi$, $\xi^{(1)}$, $\rho_1$ and $\rho_2$ are also
obtained using these wave functions. They are extracted from the
overlap
of the wave function of the ground state with the boosted wave
function
of the appropriate excited state. The boost we use is a Galilean
boost,
which means that we are neglecting relativistic effects, as well as
effects that arise from Wigner rotations. The explicit forms we obtain
for
these form factors are

\begin{eqnarray}\label{weakformfactors}
\xi(\nu)&=&\exp{\left[\frac{\bar\Lambda^2}{4\beta^2}\left(\nu^2-1
\right)
\right]},\nonumber\\
\xi^{(1)}(\nu)&=&-\sqrt{\frac{2}{3}}\left[\frac{{\bar\Lambda}^2}{4
\beta^
2}\left(\nu^2-1\right)\right]
\exp{\left[\frac{{\bar\Lambda}^2}{4\beta^2}\left(\nu^2-1\right)
\right]},
\nonumber\\
\rho_{1}(\nu)&=&\frac{1}{\sqrt{2}}\frac{\bar\Lambda}{\beta}\left(
\frac{2
\beta\beta^\prime}{\beta^2+\beta^{\prime 2}}\right)^{5/2}
\exp{\left[\frac{{\bar\Lambda}^2}{2\left(\beta^2+\beta^{\prime
2}\right)}\left(\nu^2-1\right)\right]},\nonumber\\
\rho_{2}(\nu)&=&\frac{1}{2\sqrt{2}}\left(\frac{\bar\Lambda}{\beta}
\right
)^2\left(\frac{2\beta\beta^\prime}{\beta^2+\beta^{\prime
2}}\right)^{7/2}
\exp{\left[\frac{{\bar\Lambda}^2}{2\left(\beta^2+\beta^{\prime
2}\right)}\left(\nu^2-1\right)\right]}.
\end{eqnarray}
In these expressions, $\beta$ and $\beta^\prime$ are the harmonic
oscillator parameters of the ground and excited states, respectively.
$\bar\Lambda$
is defined by writing the mass of the ground state as
$M_{(0^-,1^-)}=m_Q+\bar\Lambda$. In the second of

eqn. (\ref{weakformfactors}), we have set $\beta=\beta^\prime$ to
ensure orthogonality of the wave functions of the $(0^-,1^-)$ and
$(0^-,1^-)^\prime$
multiplets.

The parameters obtained by the methods outlined above will be refered
to as set I. We obtain a second set of values for $\beta$ and
$\beta^\prime$ (and for all of the quantities we need, except the
masses of the states) by first setting all the $\beta$'s to the same
value, and then choosing this value so that it reproduces the
experimentally measured slope of the Isgur-Wise function. The values
of
the parameters we obtain in this way are also shown in table
\ref{qmparameters}, and we will refer to this set of parameters as set
II.

\begin{table}
\caption{Quark-model parameters and low-energy constants used in this
work, sets I and II.}
\label{qmparameters}
\begin{tabular}{|c||c|c|c|c|}\hline
Multiplet & $\beta$ (GeV) & $M-M_{(0^-,1^-)}$ (GeV) & $\Gamma$ (MeV) &
coupling constant \\ \hline\hline
$(0^-,1^-)$ & 0.57 & 0 & 0 & 0.50 \\ \hline

$(0^-,1^-)^{\prime}$ & 0.57 & 0.56 & 191 & 0.69 \\ \hline

$(0^+,1^+)$ & 0.56 & 0.39 & 1040 & -1.43 \\ \hline

$(1^-,2^-)$ & 0.51 & 0.71 & 405 & -0.14 \\ \hline
\end{tabular}
\begin{tabular}{|c||c|c|c|c|}\hline
Multiplet & $\beta$ (GeV) & $M-M_{(0^-,1^-)}$ (GeV) & $\Gamma$ (MeV) &
coupling constant \\ \hline\hline
$(0^-,1^-)$ & 0.5 & 0 & 0 & 0.50 \\ \hline

$(0^-,1^-)^{\prime}$ & 0.5 & 0.56 & 174 & 0.66 \\ \hline

$(0^+,1^+)$ & 0.5 & 0.39 & 756 & -1.22 \\ \hline

$(1^-,2^-)$ & 0.5 & 0.71 & 408 & -0.215 \\ \hline
\end{tabular}
\end{table}

\section {$B_{\ell 4}$ DECAY AMPLITUDES AND  DIFFERENTIAL WIDTHS }

\subsection{$B\to D  \pi \ell \bar{\nu}  $ decay amplitude}

The $B_{\ell 4}$ decays we consider are $B^{0}\rightarrow D^{0} \ell
\bar{\nu} \pi^{+}$,
$B^{0}\rightarrow D^{+} \ell  \bar{\nu} \pi^{0}$,
$B^{-}\rightarrow D^{+} \ell  \bar{\nu} \pi^{-}$,
and $B^{-}\rightarrow D^{0} \ell  \bar{\nu} \pi^{0}$, whose amplitude
magnitudes are in the ratios $\sqrt{2}:1:\sqrt{2}:1$.
In what follows we give the results for the $\pi^{0}$ in the final
state.
The amplitude for this process has the general form
\begin{equation}\label{amplitude1}
T=\kappa\,j_{\mu} \Omega^{\mu},~~~~
\kappa\equiv V_{cb} \frac{G_{F}}{\sqrt{2}}\sqrt{M_{B}\,M_{D}}
\end{equation}
Here $j_{\mu}$ is the V-A charged leptonic current. For all practical
purposes the lepton mass can be neglected
(we do not consider decays into the $\tau$ family) and the leptonic
current is considered to be
conserved.  The hadronic part of the amplitude $ \Omega^{\mu}$
receives
non-resonant ($NR$) and resonant ($R$)
contributions,  illustrated in figures \ref{bdpienu} (a) and (b),
respectively.  Using the results of Appendices A  and B,
the evaluation of the Feynman diagrams is straightforward. The
non-resonant portion is
\begin{eqnarray}\label{nonresonant1}
\Omega_{\mu}^{NR}&=&\frac{g}{F_{0}} \xi(\nu)\,p_{\nu}\left\{\frac{
\Theta^{\nu\rho}(v)}{- 2(  p     \!\cdot\! v+\delta m_{B})+i\epsilon}
\left[\vphantom{\frac{p     \!\cdot\! v}{2(-p     \!\cdot\! v- \delta
\tilde{m}_{1} )}}i\,v^{\alpha}v^{\prime \beta}\,\epsilon_{\mu\alpha
\beta\rho}
+g_{\mu\rho}\,(1+\nu)-v_{\mu}\,v^{\prime}_{\rho}\right]\right.
\nonumber \\
&+&\left.\frac{\Theta^{\nu\rho}(v^{\prime})}{2 (p     \!\cdot\! v^{
\prime}-\delta m_{D})+i\epsilon}
\left[\vphantom{\frac{p     \!\cdot\! v}{2(-p     \!\cdot\! v- \delta
\tilde{m}_{1} )}}i\,v^{\alpha}v^{\prime \beta}\,\epsilon_{\mu\alpha
\beta\rho}
+ g_{\mu\rho}\,(1+\nu)-v_{\rho}\,v^{\prime}_{\mu}\right]\right\},
\end{eqnarray}
here $\delta m_{D}=m_{D^*}-m_{D}$ and $\delta m_{B}=m_{B^*}-m_{B}$
are the (positive) hyperfine splittings in the ground state
multiplets, and
\begin{equation}
\Theta^{\mu\nu}(v)\equiv g^{\mu\nu}-v^{\mu}v^{\nu}.
\end{equation}

The resonant portion of $ \Omega^{\mu}$ is
\begin{eqnarray}\label{resonant1}
\Omega_{\mu}^{R}&=&\frac{\alpha_{1}}{F_{0}}  \rho_{1}(\nu)\,
(v-v^{\prime})_{\mu}
\left[ -\frac{p     \!\cdot\! v}{2(-p     \!\cdot\! v- \delta
\tilde{m}_{1} )}+
 \frac{p     \!\cdot\! v^{\prime}}{2(p     \!\cdot\! v^{\prime}-
\delta \tilde{m}_{1} )}\right]\nonumber\\
&+&\frac{\alpha_{2}}{3F_{0}}  \rho_{2}(\nu)\,p_{\rho}\,\left\{
\frac{\Theta^{\nu\rho}(v)}{2(-p     \!\cdot\! v- \delta \tilde{m}_{2}
)}
\left[\vphantom{\frac{p     \!\cdot\! v}{2(-p     \!\cdot\! v- \delta
\tilde{m}_{1} )}}\,
i\epsilon_{\mu\nu\alpha\beta}\,v^{\alpha}v^{\prime\beta}\,( \nu-1)
\right.\right.\nonumber\\
&+&\left.\left.\vphantom{\frac{p     \!\cdot\! v}{2(-p     \!\cdot\!
v- \delta \tilde{m}_{1} )}}g_{\mu\nu} (\nu^{2}-1)-v^{\prime}_{\nu}
\left[v_{\mu}(2+\nu)
-3v^{\prime}_{\mu}\right]\right]\right.\nonumber\\
&+&\left.\frac{\Theta^{\nu\rho}(v^{\prime})}{2(p     \!\cdot\! v^{
\prime}- \delta \tilde{m}_{2} )}
\left[\vphantom{\frac{p     \!\cdot\! v}{2(-p     \!\cdot\! v- \delta
\tilde{m}_{1} )}}+i\epsilon_{\mu\nu\alpha\beta}v^{\alpha}v^{\prime
\beta}(\nu-1)+
g_{\mu\nu}( \nu^{2}-1)\right.\right.\nonumber\\
&-&\left.\left.v_{\nu}\left[v^{\prime}_{\mu}(2+\nu)-3v_{\mu}\right]
\vphantom{\frac{p     \!\cdot\! v}{2(-p     \!\cdot\! v- \delta
\tilde{m}_{1} )}}\right]\right\}\nonumber\\
&+&\frac{\alpha_{3}}{F_{0}} \xi^{(1)}(\nu)\,p_{\nu}
\left\{\frac{\Theta^{\nu\rho}(v)}{- 2  p     \!\cdot\! v- \delta
\tilde{m}_{3} }
\left[\vphantom{\frac{p     \!\cdot\! v}{2(-p     \!\cdot\! v- \delta
\tilde{m}_{1} )}}i\,v^{\alpha}v^{\prime \beta}\,\epsilon_{\mu\alpha
\beta\rho}
+g_{\mu\rho}\,(1+\nu)-v_{\mu}\,v^{\prime}_{\rho}\right]\right.
\nonumber\\
&+&\left.\frac{\Theta^{\nu\rho}(v^{\prime})}{2 p     \!\cdot\! v^{
\prime}- \delta \tilde{m}_{3}}
\left[\vphantom{\frac{p     \!\cdot\! v}{2(-p     \!\cdot\! v- \delta
\tilde{m}_{1} )}}i\,v^{\alpha}v^{\prime \beta}\,\epsilon_{\mu\alpha
\beta\rho}
+ g_{\mu\rho}\,(1+\nu)-v_{\rho}\,v^{\prime}_{\mu}\right]\right\}.
\end{eqnarray}
Here we denote $\delta \tilde{m}_{j} \equiv \delta {m}_{j}-i
\Gamma_{j}/2$, where $ \Gamma_{j}$  is the total width of the
resonance.

As explained earlier, contributions from other resonances than the
ones considered are
 suppressed either by higher
powers of the soft-pion momentum, as is the case with the
$(1^{+},2^{+})$
multiplet,  by higher powers of $( \nu-1)$ in the
small recoil domain, or by the fact that they are much heavier than
the ground state mesons.

\subsection{$B\rightarrow D^{\ast}  \pi\ell \bar{\nu} $ decay
amplitude}

The amplitude  for this process has the general form
\begin{equation}\label{amplitude2}
T=\kappa j_{\mu} \Omega^{\mu\nu}\epsilon_{D^{\ast}\nu},
\end{equation}
where $\epsilon_{D^{\ast}\nu}$ is the polarization vector of the $D^{
\ast}$.

The non-resonant contributions to $ \Omega^{\mu\nu}$  are obtained
from the
diagrams shown in figure \ref{bdstarpienu} (a), which give
\begin{eqnarray} \label{nonresonant2}
 \Omega_{\mu\nu}^{NR}&=&\frac{g}{2\,F_{0}} \xi(\nu)\;\left\{
\vphantom{\frac{\Theta_{\rho\nu}(v)}{-p     \!\cdot\! v+i\epsilon}}
(v+v^{\prime})_{\mu} \frac{p_{\nu}}{p     \!\cdot\! v^{\prime}+\delta
m_{D}+i\epsilon}\right.\nonumber\\
&+&p^{\rho}\frac{\Theta_{\rho\sigma}(v)}{-(p     \!\cdot\! v+\delta
m_{B})+i\epsilon}
\left[\vphantom{\frac{\Theta_{\rho\nu}(v)}{-(p     \!\cdot\! v+\delta
m^{\ast}_{B})+i\epsilon}}
g_{\nu\sigma}(v+v^{\prime})_{\mu}
-g_{\mu\sigma} v_{\nu}-g_{\mu\nu}v^{\prime}_{\sigma}+
i\epsilon_{\mu\alpha\sigma\nu} (v+v^{\prime})^{\alpha}\right]
\nonumber\\
&+&\left.\frac{\Theta_{\rho\delta}(v^{\prime})}{p     \!\cdot\! v^{
\prime}+i\epsilon}
 \left[\vphantom{\frac{\Theta_{\rho\delta}(v)}{-(p     \!\cdot\! v+
\delta m^{\ast}_{B})+i\epsilon}}
-i\epsilon_{\mu\rho\alpha\beta}v^{\prime\alpha}v^{\beta}+
g_{\mu\rho}(1+\nu)-v_{\rho}v^{\prime}_{\mu}\right]\,i\epsilon_{\omega
\nu\gamma\delta}
p^{\gamma}v^{\prime\omega}\right\}.
\end{eqnarray}

The resonant contributions obtained from the diagrams  in figure
\ref{bdstarpienu} (b) are given by
\begin{eqnarray} \label{resonant2}
 \Omega_{\mu\nu}^{R}&=&
\frac{\alpha_{1}}{F_{0}}  \rho_{1}(\nu)\;\left[\vphantom{\frac{
\Theta_{\rho\nu}(v)}{-2p     \!\cdot\! v+i\epsilon}}
-g_{\mu\nu}(\nu-1)+v^{\prime}_{\mu}v_{\nu}-i\epsilon_{\mu\nu\alpha
\beta}
v^{\alpha}v^{\prime\beta}\right]\nonumber\\
&\times&\left[-\frac{p     \!\cdot\! v}{2(-p     \!\cdot\! v- \delta
\tilde{m}_{1} )}+
\frac{p     \!\cdot\! v^{\prime}}{2(p     \!\cdot\! v^{\prime}-
\delta \tilde{m}_{1} )}\right]\nonumber\\
&-&\frac{\alpha_{2}}{3F_{0}}  \rho_{2}(\nu)\;\left\{
p_{\rho}\frac{\Theta^{\rho\sigma}(v)}{2(-p     \!\cdot\! v- \delta
\tilde{m}_{2} )}
\left[\vphantom{\frac{\Theta_{\rho\nu}(v)}{-2p     \!\cdot\! v+i
\epsilon}}
-g_{\nu\sigma}(v+v^{\prime})_{\mu}( \nu-1)+
3v_{\nu}v^{\prime}_{\mu} v^{\prime}_{\sigma}\right.\right.\nonumber\\
&-&2 g_{\mu\nu} v^{\prime}_{\sigma}(\nu-1)+
g_{\mu\sigma} v_{\nu} (\nu-1)+i\epsilon_{\mu\sigma\alpha\nu}
(v-v^{\prime})^{\alpha} (1+\nu)\nonumber\\
&+&\left.\left.\vphantom{\frac{\Theta_{\rho\nu}(v)}{-2p     \!\cdot\!
v+i\epsilon}}
\,2i\epsilon_{\sigma\nu\alpha\beta} v^{\prime\alpha}v^{\beta}
v^{\prime}_{\mu}+i\epsilon_{\mu\nu\alpha\beta}
v^{\prime}_{\sigma}v^{\prime \alpha}v^{\beta}\right]\right.\nonumber\\
&+& \left. 3 p^{\gamma} \frac{\Theta^{\delta\rho}_{\gamma\nu}(v^{
\prime})}
{2(p     \!\cdot\! v^{\prime}- \delta \tilde{m}_{2} )+i\epsilon}
\left[\vphantom{\frac{\Theta_{\rho\nu}(v)}{-2p     \!\cdot\! v+i
\epsilon}}
v_{\delta} g_{\rho\mu}(\nu-1)-v_{\delta}v_{\rho}v^{\prime}_{\mu}+
i\epsilon_{\mu\delta\alpha\beta}v_{\rho} v^{\alpha}
 v^{\prime \beta}\right] \right.\nonumber\\
&+&  \left.\frac{i}{2}\frac{\Theta^{\sigma\rho}(v^{\prime})}{2 (p
\!\cdot\! v^{\prime}- \delta \tilde{m}_{2} )}
\epsilon_{\rho\delta\gamma\nu} p_{\pi}^{\delta} v^{\prime\gamma}
\left[\vphantom{\frac{\Theta_{\rho\nu}(v)}{2(-p     \!\cdot\! v-
\delta \tilde{m}_{3} )}}\,i \epsilon_{\mu\sigma\alpha\beta}v^{
\alpha}v^{\prime\beta} (\nu-1)+
g_{\mu\sigma} (\nu^{2}-1)\right.\right.\nonumber\\
&-&\left.\left.v_{\sigma}\left[v^{\prime}_{\mu}(2+\nu)-3v_{\mu}\right]
\vphantom{\frac{\Theta_{\rho\nu}(v)}{2(-p     \!\cdot\! v- \delta
\tilde{m}_{3} )}}\right]\right\}\nonumber\\
&+&\frac{\alpha_{3}}{F_{0}} \xi^{(1)}(\nu)\;\left\{\vphantom{\frac{
\Theta_{\rho\nu}(v)}{2(-p     \!\cdot\! v- \delta \tilde{m}_{3} )}}
(v+v^{\prime})_{\mu} \frac{p_{\nu}}{2(p     \!\cdot\! v^{\prime}-
\delta \tilde{m}_{3} )}\right.\nonumber\\
&+&\left.p^{\rho}\frac{\Theta_{\rho\sigma}(v)}{2(-p     \!\cdot\! v-
\delta \tilde{m}_{3} )}
\left[\vphantom{\frac{\Theta_{\rho\nu}(v)}{2(-p     \!\cdot\! v-
\delta \tilde{m}_{3} )}}
g_{\nu\sigma}(v+v^{\prime})_{\mu}
-g_{\mu\sigma} v_{\nu}-g_{\mu\nu}v^{\prime}_{\sigma}+
i\epsilon_{\mu\alpha\sigma\nu} (v+v^{\prime})^{\alpha}\right]\right.
\nonumber\\
&+&\left.\frac{\Theta_{\rho\delta}(v^{\prime})}{2(p     \!\cdot\! v^{
\prime}- \delta \tilde{m}_{3} )}
 \left[\vphantom{\frac{\Theta_{\rho\delta}(v)}{2(-p     \!\cdot\! v-
\delta \tilde{m}_{3} )}}
-i\epsilon_{\mu\rho\alpha\beta}v^{\prime\alpha}v^{\beta}+
g_{\mu\rho}(1+\nu)-v_{\rho}v^{\prime}_{\mu}\right]\,i\epsilon_{\omega
\nu\gamma\delta}
p^{\gamma}v^{\prime\omega}\right\}.
\end{eqnarray}
Here, $\Theta^{\delta\rho}_{\gamma\nu}$ results from the numerator
of the spin-2 propagator in the heavy mass limit, and is
\begin{equation}\label{thetamunurholam}
\Theta^{\mu\nu}_{\rho\sigma}(v)=\frac{1}{2} \Theta^{\mu}_{\rho}
\Theta^{\nu}_{\sigma}+
\frac{1}{2} \Theta^{\mu}_{\sigma} \Theta^{\nu}_{\rho}-\frac{1}{3}
\Theta^{\mu\nu} \Theta_{\rho\sigma}.
\end{equation}
As expected, all amplitudes vanish in the soft-pion limit. Moreover,
resonance contributions vanish at zero recoil, as predicted by the
heavy mass limit.

\subsection{$B\to D \pi \ell \bar{\nu}  $ decay rate}

In the analysis of $B_{\ell 4}$ decays it is convenient
to use  the momentum combinations
\begin{eqnarray}\label{momenta}
P&=&p_{D}+p_{\pi},~~~~~~~~~~p_{D}=M_{D} v^{\prime},~~~~p_{\pi}=p,
\nonumber\\
Q&=&p_{D}-p_{\pi},\nonumber\\
L&=&p_{\ell}+p_{\nu},\nonumber\\
N&=&p_{\ell}-p_{\nu}.
\end{eqnarray}
In terms of these variables, the most general form of
$\Omega_{\mu}$ is
\begin{equation}\label{omega1}
\Omega_{\mu}=\frac{i}{2} H \,\epsilon_{\mu\nu\rho\sigma} L^{\nu}Q^{
\rho}P^{\sigma}+
FP_{\mu}+G Q_{\mu}+R L_{\mu}
\end{equation}
where $H$, $F$, $G$ and $R$ are form factors dependent on the
three invariants $\nu$, $p     \!\cdot\! v$ and $p     \!\cdot\! v^{
\prime}$. These form factors
are easily obtained from the explicit expressions of the
non-resonant and resonant parts of the amplitude given in eqns. (
\ref{nonresonant1}) and (\ref{resonant1}).
The explicit expressions for the form factors are given in Appendix C.
The assumption that the leptonic current is conserved implies that the
term proportional to $R$ does not contribute and can be ignored.

\begin{figure}
\let\picnaturalsize=N
\def\picsize{2.5in}
\def\picfilenamea{bdpienu_angles.ps}
\ifx\nopictures Y\else{\ifx\epsfloaded Y\else\input epsf \fi
\let\epsfloaded=Y
\centerline{
\ifx\picnaturalsize N\epsfxsize \picsize\fi \epsfbox{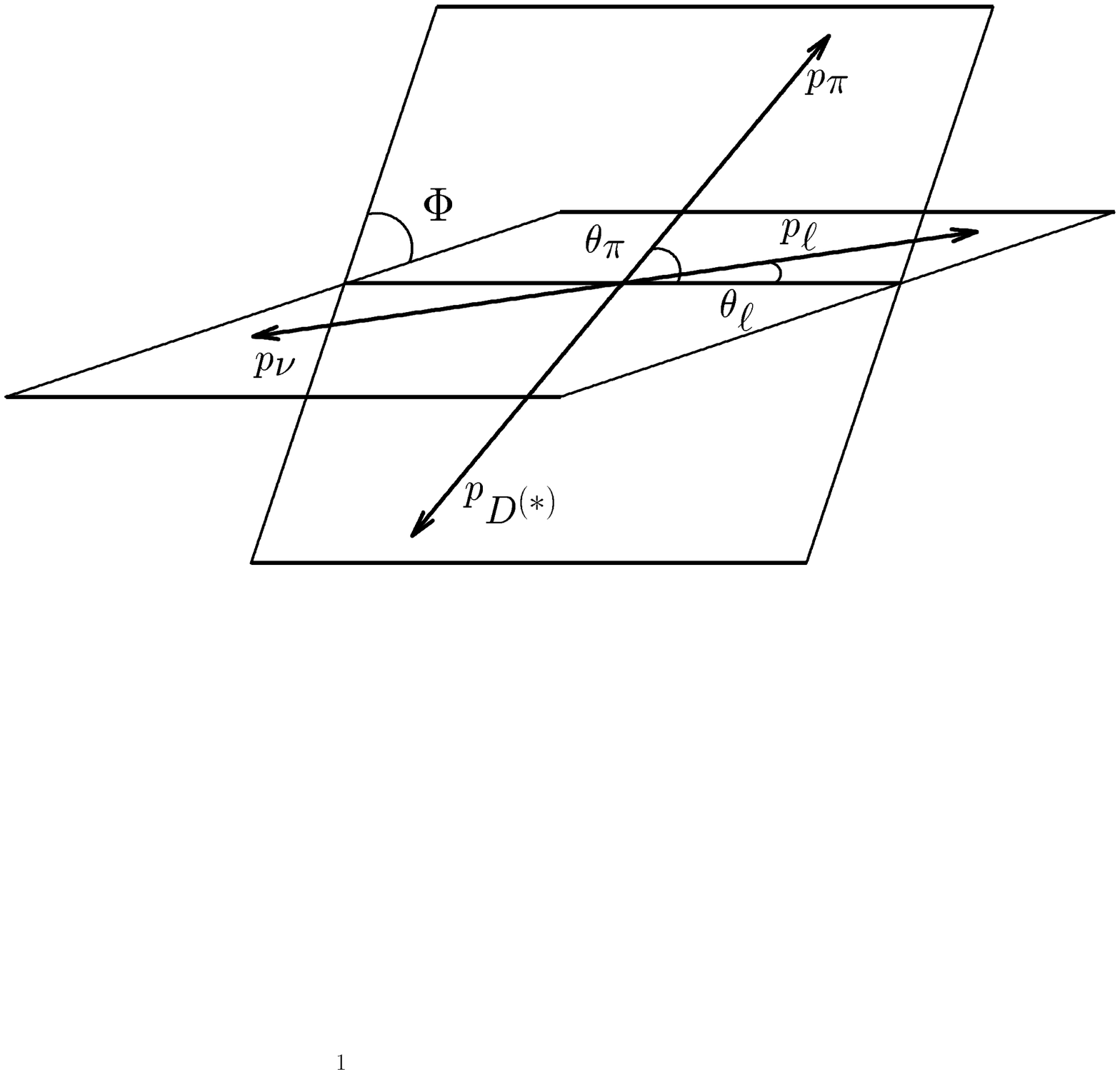}}}
\fi
\vspace*{-3.0cm}
\caption{Kinematic variables and angles. \label{kinematics}}
\end{figure}

The squared modulus of the decay amplitude, after summing over the
lepton polarizations
and neglecting higher order terms in the pion mass squared is given by
\begin{eqnarray}\label{tsquared1}
\sum_{spins} |T|^{2}&=&\kappa^{2}\;\left\{|F|^{2}\left[
\lambda(M_{B}^{2},S_{D \pi},S_{\ell})-
4 \,(N     \!\cdot\! P)^{2}\right]\right.\nonumber\\
&+&|G|^{2}\left[4 \,(L     \!\cdot\! Q)^{2}+4 \,S_{\ell}
 (S_{D \pi}-2 M_{D}^{2})-4 \,(N     \!\cdot\! Q)^{2}\right]\nonumber\\
&+&|H|^{2}\left[-M_{D}^{4} \,S_{\ell}^{2}-\frac{1}{4}\, S_{
\ell}(2M_{D}^{2}-S_{D \pi})
(M_{B}^{2}-S_{\ell}-S_{D \pi})^{2}\right.\nonumber\\
&-&(L     \!\cdot\! Q)^{2} \,S_{\ell}S_{D \pi}
 + L     \!\cdot\! Q\, M_{D}^{2}\, S_{\ell}
(M_{B}^{2}-S_{\ell}-S_{D \pi})\nonumber\\
&+& S_{\ell}^{2} \,S_{D \pi} (2 \,M_{D}^{2}- S_{D \pi})
-\left.\vphantom{\frac{1}{4}}\,(\epsilon_{\mu\nu\rho\sigma} L^{
\mu}N^{\nu}P^{\rho}Q^{\sigma})^{2}
\right] \nonumber\\
&+&4 \;{\rm Re}(FG^{\ast})\left[ -2\, M_{D}^{2}\, S_{\ell}+L     \!
\cdot\! Q (M_{B}^{2}-S_{\ell}-S_{D \pi})
-2\, N     \!\cdot\! P N     \!\cdot\! Q\right]\nonumber\\
&+& {\rm Re}(FH^{\ast})\left[4\, M_{D}^{2}\, S_{\ell} N     \!\cdot\!
P-2 \,L     \!\cdot\! Q\; N     \!\cdot\! P
(M_{B}^{2}-S_{\ell}-S_{D \pi})\right.\nonumber\\
&+&\left.N     \!\cdot\! Q \;\lambda(M_B^2, S_{D \pi}, S_{\ell})
\right]\nonumber\\
&+& {\rm Re}(G H^{\ast})\left[4\, S_{\ell} N     \!\cdot\! P
(2M_{D}^{2}-S_{D \pi})
-4\, (L     \!\cdot\! Q)^{2} N     \!\cdot\! P-4\, M_{D}^{2} S_{\ell}
N     \!\cdot\! Q\right.\nonumber\\
&+&\left. 2 \,L     \!\cdot\! Q\;N     \!\cdot\! Q (M_{B}^{2}-S_{
\ell}-S_{D \pi})\right]\nonumber\\
&+&\left. 4\; {\rm Im}(2\, F^{\ast} G + F^{\ast} H N     \!\cdot\!
P+G^{\ast}H N     \!\cdot\! Q)
\;\epsilon_{\mu\nu\rho\sigma} L^{\mu}N^{\nu}P^{\rho}Q^{\sigma}\right\}
\end{eqnarray}
The invariants appearing in this expression are
\begin{eqnarray} \label{invariants1}
P^{2}&=&S_{D \pi},\nonumber\\
L^{2}&=&-N^{2}= S_{\ell},\nonumber\\
P     \!\cdot\! Q&=&M_{D}^{2}-M_{\pi}^{2},\nonumber\\
P     \!\cdot\! L&=&\frac{1}{2} \,(M_{B}^{2}-S_{\ell}-S_{D \pi}),
\nonumber\\
L     \!\cdot\! N&=&0.
\end{eqnarray}
and  $\lambda$ is  K\"all\`en's function.
In order to obtain explicit expressions for the remaining
invariants, namely $Q^{2}$, $P     \!\cdot\! N$, $N     \!\cdot\! Q$
and $L     \!\cdot\! Q$,
it is convenient to use as independent variables the quantities $S_{D
\pi}$, $S_{\ell}$,
and the angles $\theta_{\pi}$, $\theta_{\ell}$ and $\phi$.
{}From eqn. (\ref{invariants1}), $S_{D \pi}$ and $S_{\ell}$ are the
invariant mass squared of the $\pi D$ and $\ell{\nu}$ pairs,
respectively.
The angles $\theta_{\pi}$, $\theta_{\ell}$ and $\phi$ are illustrated
in figure \ref{kinematics}.
$\theta_{\pi}$ is the angle between the pion momentum
and the direction of $\overrightarrow{P}$ in the c.m. frame of the
$\pi D$ pair, $\theta_{\ell}$ is the angle between the lepton
momentum and the direction of $\overrightarrow{L}$ in the c.m. frame
of the
$\ell\bar{\nu}$ pair, and $\phi$  is the angle between the two decay
planes
defined by the  pairs $(\overrightarrow{p_{\pi}},
\overrightarrow{p_{D}})$  and
 $(\overrightarrow{p_{\ell}},\overrightarrow{p_{\nu}})$ in the rest
frame of the $B$-meson.
This is the set of variables initially introduced by Cabibbo
and Maksymowicz \cite{cabibbo} in the analysis of $K_{\ell 4}$ decays.

The remaining invariants, are
\begin{eqnarray} \label{invariants2}
Q^{2}&=&-\frac{1}{4 M_{B}^{2} S_{D \pi}^{2}}\;\left[(M_{D}^{2}-
M_{\pi}^{2})\;\lambda^{1/2}(M_{B}^{2},S_{D \pi}, S_{\ell})\right.
\nonumber\\
&+&\left.\cos\theta_{\pi} (M_{B}^{2}+S_{D \pi}-S_{\ell})\;
\lambda^{1/2}(M_{D}^{2},S_{D \pi},M_{\pi}^{2})\right]^{2}\nonumber\\
&+&\frac{1}{4 M_{B}^{2} S_{D \pi}^{2}}\;\left[
(M_{D}^{2}-M_{\pi}^{2}) (M_{B}^{2}+S_{D \pi}-S_{\ell})\right.
\nonumber\\
&+&\left.\cos\theta_{\pi}\;\lambda^{1/2}(M_{B}^{2},S_{D \pi}, S_{
\ell})\;
\lambda^{1/2}(M_{D}^{2},S_{D \pi},M_{\pi}^{2})\right]^{2}\nonumber\\
&-&\frac{1}{S_{D \pi}}\,\sin^{2}\theta_{\pi} \;
\lambda(M_{D}^{2},S_{D \pi},M_{\pi}^{2}),\nonumber\\
P     \!\cdot\! N&=&\frac{1}{2} \cos\theta_{\ell} \;
\lambda^{1/2}(M_{B}^{2},S_{D \pi}, S_{\ell}),\nonumber\\
L     \!\cdot\! Q&=&\frac{1}{2 S_{D \pi}}
\left[M_{D}^{2}(M_{B}^{2}-S_{D \pi}-S_{\ell})\right.\nonumber\\
&+&\left.\cos\theta_{\pi}\;\lambda^{1/2}(M_{D}^{2},S_{D \pi},M_{
\pi}^{2})\;
\lambda^{1/2}(M_{B}^{2},S_{D \pi}, S_{\ell})\right],
\nonumber\\
N     \!\cdot\! Q&=&\frac{M_{D}^{2}}{2 S_{D \pi}}
\cos\theta_{\ell}\;\lambda^{1/2}(M_{B}^{2},S_{D \pi}, S_{\ell})
\nonumber\\
&+&\frac{1}{4 M_{B}^{2} S_{D \pi}}\cos\theta_{\ell}\cos\theta_{\pi}\;
\lambda^{1/2}(M_{D}^{2},S_{D \pi},M_{\pi}^{2})\nonumber\\
&\times&\left(S_{\ell}S_{D \pi}+M_{B}^{4}-S_{\ell}^{2}-S_{D \pi}^{2}+
\lambda(M_{B}^{2},S_{D \pi}, S_{\ell})\right)\nonumber\\
&-&\sqrt{\frac{S_{\ell}}{S_{D \pi}}} \cos\phi \sin\theta_{\ell} \sin
\theta_{\pi}\;
\lambda^{1/2}(M_{D}^{2},S_{D \pi},M_{\pi}^{2}),\nonumber\\
\epsilon_{\mu\nu\rho\sigma} P^{\mu}Q^{\nu}L^{\rho}N^{\sigma}&=&
-\frac{1}{2} \sqrt{\frac{S_{\ell}}{S_{D \pi}}}\;
\lambda^{1/2}(M_{D}^{2},S_{D \pi},M_{\pi}^{2})\;
\lambda^{1/2}(M_{B}^{2},S_{D \pi}, S_{\ell})\nonumber\\
&\times&\sin\phi
 \sin\theta_{\ell} \sin\theta_{\pi}.
\end{eqnarray}
Since the form factors depend on only one of the angles, namely
$\theta_{\pi}$,  in the expression for the partial width
of the $B_{\ell 4}$ decay the integrations over the angles $\phi$ and
$\theta_{\ell}$ can be performed explicitly.
Following standard steps, the differential partial width of interest
can be expressed as
\begin{eqnarray}\label{dgamma}
&&\frac{d^{3}\Gamma_{B_{\ell 4}}}{d\cos\theta_{\pi}dS_{D \pi}dS_{
\ell}}=
\frac{\aleph}{2 M_{B}} J(S_{D \pi}, S_{\ell}) \,
\int_{0}^{2\pi} d\phi\int_{-1}^{1} d\cos\theta_{\ell} \nonumber\\
&\times&\sum_{spins}
|T|^{2}(S_{D \pi}, S_{\ell}, \theta_{\pi}, \theta_{\ell}, \phi),
\end{eqnarray}
where
\begin{equation}
\aleph=\cases{2,&for charged pions\cr
1,&for neutral pions},
\end{equation}
and the Jacobian $J(x,y)$ is
\begin{equation}\label{jacobian}
J(x,y)=\frac{1}{2^{14}\pi^{6} x y M_{B}^{2}} \;
\lambda^{1/2}(M_{B}^{2},x,y)
\; \lambda^{1/2}(M_{D}^{2},M_{\pi}^{2},x)\;
\lambda^{1/2}(0,0,y).
\end{equation}
For our purposes the interesting differential partial rate
is $d\Gamma_{B_{\ell 4}}/dS_{D\pi}$ which results from integrating
eqn.
(\ref{dgamma}) over $S_{\ell}$ and $\theta_{\pi}$ with no kinematic
cuts.

\subsection{$B\rightarrow D^{\ast} \pi \ell \bar{\nu} $ decay rate}

The  tensor $\Omega_{\mu\nu}$ can be expressed in terms of twelve
form factors as
\begin{eqnarray}\label{omegamunu1}
\Omega_{\mu\nu}&=&\frac{i}{2} H_{1}\, \epsilon_{\mu\nu\rho\sigma} P^{
\rho} Q^{\sigma}+
\frac{i}{2} H_{2}\, \epsilon_{\mu\nu\rho\sigma}P^{\rho} L^{\sigma}+
\frac{i}{2} H_{3}\, \epsilon_{\mu\nu\rho\sigma}Q^{\rho} L^{\sigma}
\nonumber\\
&+&F_{1} P_{\mu} (P-Q)_{\nu}+F_{2} Q_{\mu}  (P-Q)_{\nu}+F_{3}P_{\mu}
L_{\nu}+
F_{4}  Q_{\mu}  L_{\nu}+ K g_{\mu\nu}\nonumber\\
&+& \frac{i}{2} G_{1}^{A} \,   \epsilon_{\mu\delta\rho\sigma}
P^{\delta}Q^{\rho}L^{\sigma} (P-Q)_{\nu} +
\frac{i}{2} G_{2}^{A}  \,  \epsilon_{ \mu  \delta\rho\sigma}
P^{\delta}Q^{\rho}L^{\sigma} L_{\nu}\nonumber\\
&+&
\frac{i}{2} G_{1}^{B}  \,  \epsilon_{\nu \delta\rho\sigma}
P^{\delta}Q^{\rho}L^{\sigma} P_{\mu} +
\frac{i}{2} G_{2}^{B} \,  \epsilon_{ \nu \delta\rho\sigma}
P^{\delta}Q^{\rho}L^{\sigma}  Q_{\mu}.
\end{eqnarray}
In writing this form, we have neglected terms that vanish upon
contraction with the conserved
leptonic current $j_{\mu}$ and with the $D^{\ast}$ polarization vector
$\epsilon_{D^{\ast}}^{\nu}$.
The explicit expressions for the form factors resulting from eqn. (
\ref{omegamunu1}) are given in Appendix C.

It is straightforward to calculate the modulus squared of the
resulting decay amplitude summed over
the polarizations of the $D^{\ast}$.
 Since the result is lengthy we prefer not to
display it here.  The partial width is given by an expression similar
to that of eqn. (\ref{dgamma})
with the appropriate replacement of the squared amplitude.

\section{RESULTS, DISCUSSION AND CONCLUSIONS}

If we throw caution to the wind and apply our calculation to all of
phase space, we find that the decay rate for
$B\to D\pi\ell\nu$ ranges from $1.09 \times 10^{-14}$ GeV to $1.13
\times 10^{-14}$ GeV. The upper limit corresponds to including all the
multiplets in the calculation, while the lower limit arises from
including only the $(0^-,1^-)$ and $(0^+,1^+)$
multiplets. These decay rates correspond
to branching fractions of 2.1\%, somewhat larger than, but largely in
agreement with the analysis of Cheng and collaborators
\cite{chengetal}. We see,
therefore, that the inclusion of the higher multiplets does not
profoundly affect the total decay rate of $B\to D\pi\ell\nu$.
The effects on the spectrum, and on the decay $B\to D^*\pi\ell\nu$ are
somewhat more striking, however.

Performing the same integration for the decay $B\to D^*\pi\ell\nu$, we
find that the total rate varies between $2.7 \times 10^{-16}$ GeV
($BR=5.0 \times 10^{-4}$) and $1.7 \times 10^{-15}$ GeV ($BR=0.3\%$).
Thus, inclusion of the resonances makes a significant change to this
decay rate, increasing it by a factor of about 6. In both cases ($B\to
D\pi\ell\nu$ and $B\to D^*\pi\ell\nu$), the change in the quark model
parameters from set I to set II has little effect on the integrated
decay rates, but makes significant differences to the spectra
obtained.

In a series of figures we show the differential $B_{\ell 4}$ decay
rates into a charged pion $\frac{\partial\Gamma_{B_{\ell 4}}}{\partial
S_{D\pi}}$ as a
function of $S_{D\pi}$, for values of $S_{D\pi}$ between the threshold
of $(M_D^{(\ast)}+m_\pi)^2$ and 10 GeV$^2$. This covers a bit beyond
the domain where the soft pion limit may be safely applied; for
$\Lambda_{\chi}\sim 0.5$ GeV, the soft pion limit holds up to
$S_{D\pi}\sim 6.5$ ${\rm GeV}^{2}$. In each of these figures, we show
the spectra resulting from both sets of parameters, with set I
corresponding to figures N(a), and set II to figures N(b). In
addition,
we normalize by dividing the differential decay width by the total
semileptonic decay width $B\to D\ell\nu$, calculated in the same
model.
In this way, we can lessen the impact of model dependences that enter
through form factors and coupling constants. We note, however, that
using $|V_{cb}|=0.043$, we obtain branching fractions of 1.8\% (1.7\%)
(here, and in all that follows, the first number is obtained using the
parameters of set I, while the second number, in parantheses, is
obtained using the parameters of set II) for $B\to D\ell\nu$ and 4.6\%
(4.6\%) for $B\to D^*\ell\nu$, in surprisingly good agreement with
experiment. We have not compared our results with the differential
decay rates for these decays, however.

\begin{figure}
\let\picnaturalsize=N
\def\picsize{2in}
\def\picfilenamea{fig5a.ps}
\def\picfilenameb{fig5b.ps}
\ifx\nopictures Y\else{\ifx\epsfloaded Y\else\input epsf \fi
\let\epsfloaded=Y
\centerline{
\ifx\picnaturalsize N\epsfxsize \picsize\fi \epsfbox{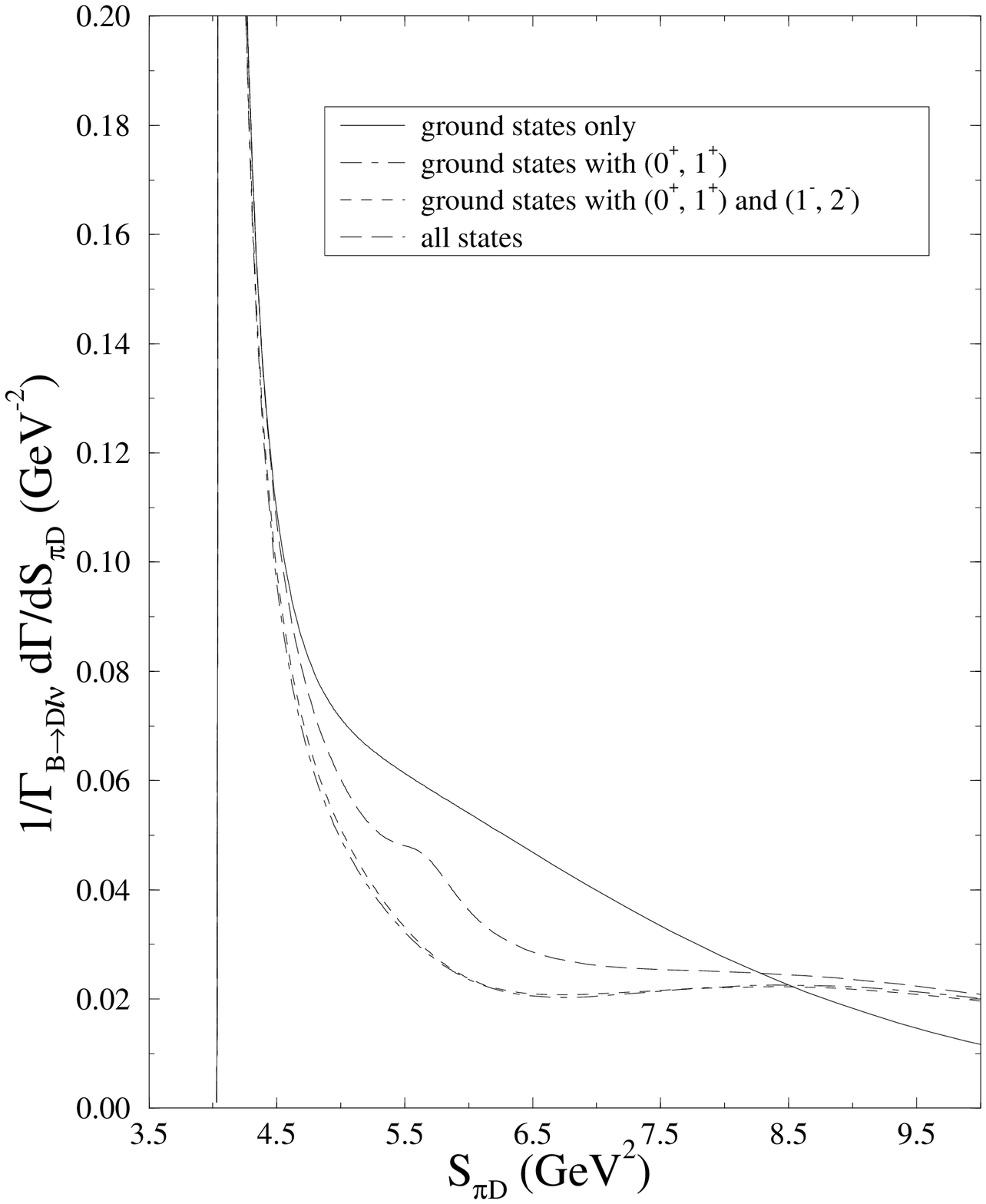}
\hfil
\ifx\picnaturalsize N\epsfxsize \picsize\fi \epsfbox{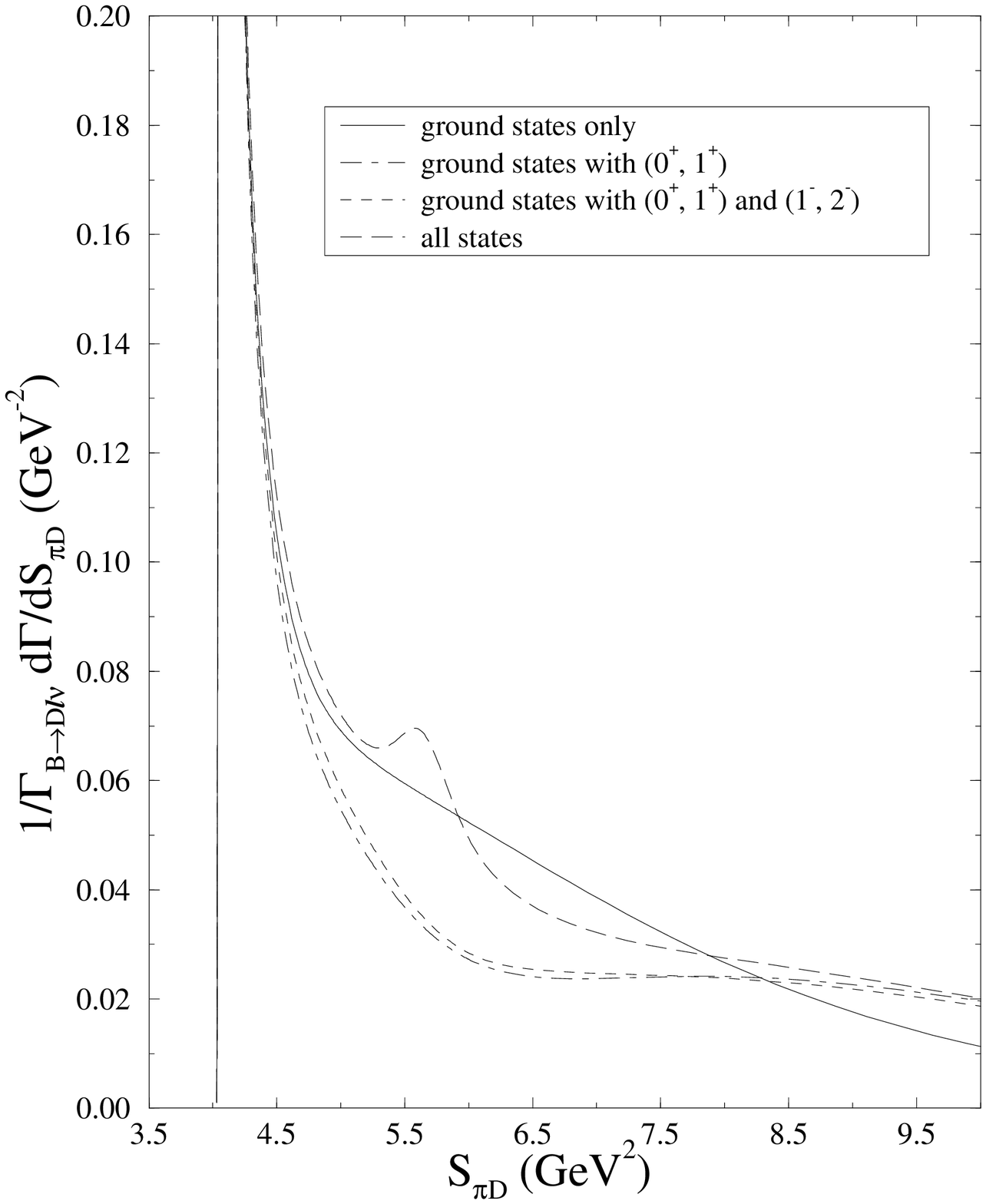}
}}\fi
\caption{$\frac{1}{\Gamma(B\rightarrow D\ell\nu)}\,\frac{\partial
\Gamma_{B_{\ell 4}}}{\partial S_{D\pi}}$ as a function of
$S_{D\pi}$ for $B\to D\pi^{\pm} e\nu$. The different curves correspond
to different combinations of resonances, as explained in
the figure.\label{spectrum1}}
\end{figure}

\begin{figure}
\let\picnaturalsize=N
\def\picsize{2in}
\def\picfilenamea{fig6a.ps}
\def\picfilenameb{fig6b.ps}
\ifx\nopictures Y\else{\ifx\epsfloaded Y\else\input epsf \fi
\let\epsfloaded=Y
\centerline{
\ifx\picnaturalsize N\epsfxsize \picsize\fi \epsfbox{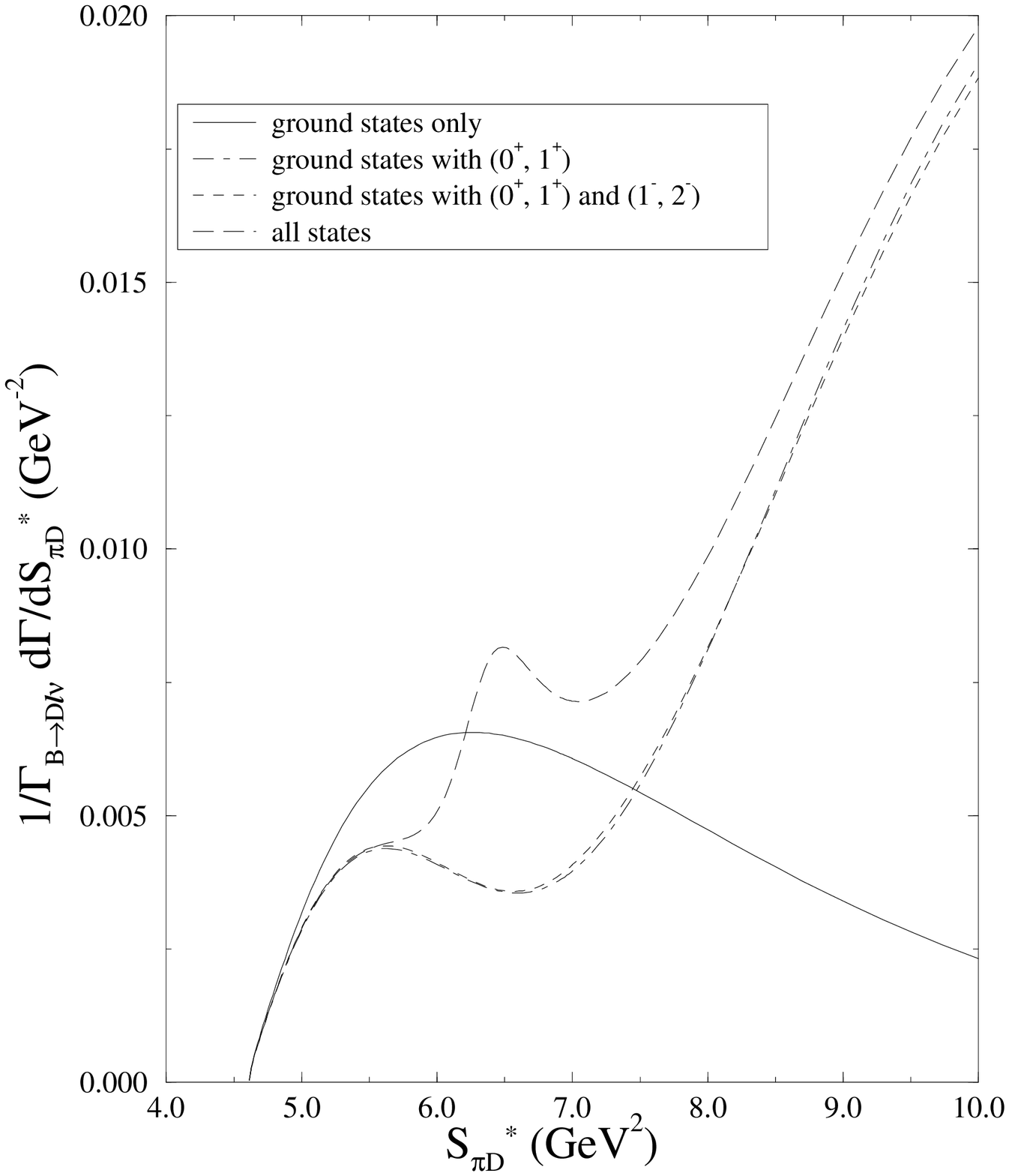}
\hfil
\ifx\picnaturalsize N\epsfxsize \picsize\fi \epsfbox{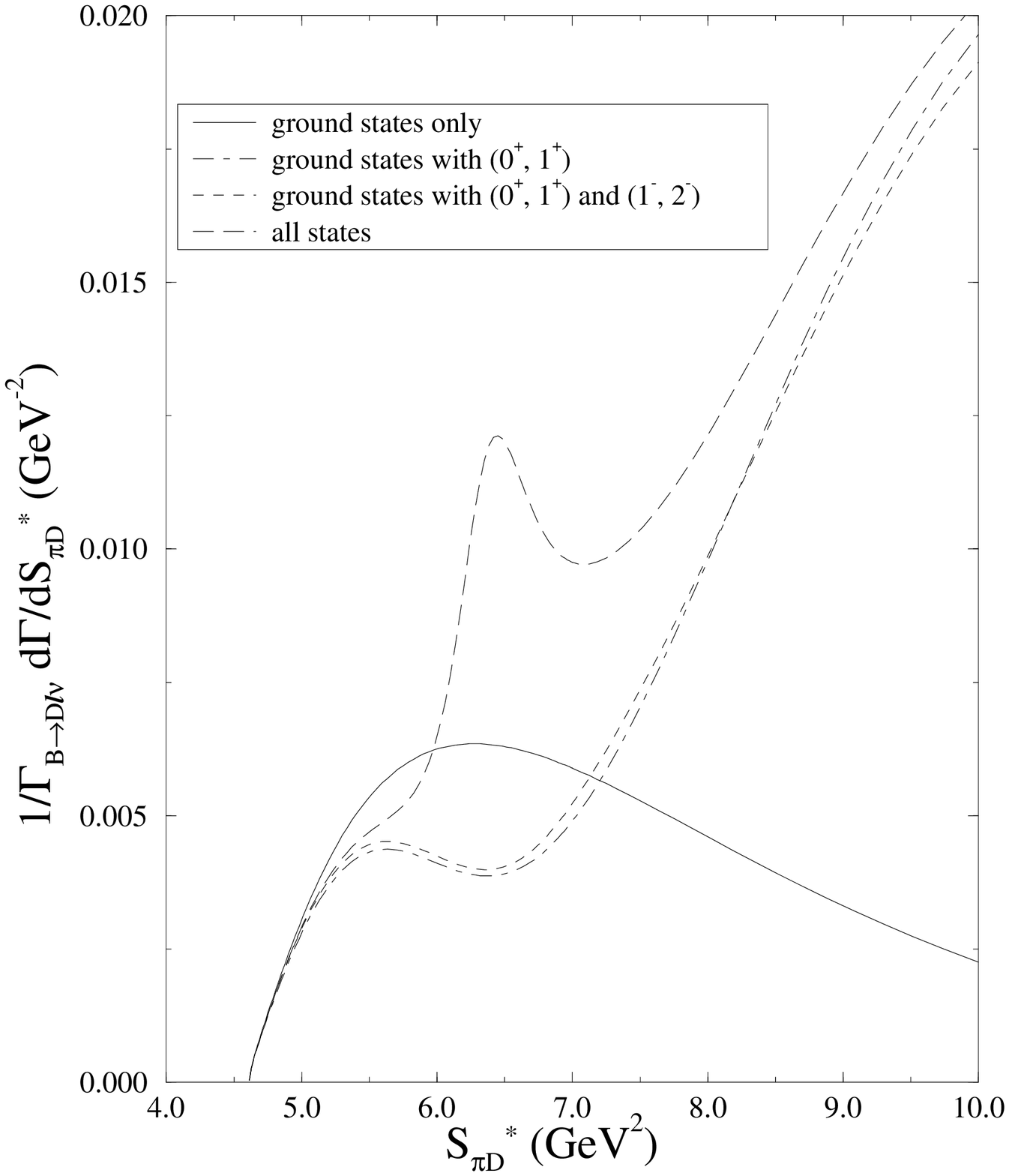}
}}\fi
\caption{$\frac{1}{\Gamma(B\rightarrow D\ell\nu)}\,\frac{\partial
\Gamma_{B_{\ell 4}}}{\partial S_{D\pi}}$ as a function of
$S_{D\pi}$ for $B\to D^{*}\pi^{\pm}  e\nu$. \label{spectrum2}}
\end{figure}

Figures \ref{spectrum1} and \ref{spectrum2} show the spectra for the
decays $B\to D\pi\ell\nu$ and $B\to D^*\pi\ell\nu$, respectively. Each
of these figures shows four curves, corresponding to the inclusion of
various combinations of multiplets in the analysis. One very
interesting feature in these spectra is the narrow structure due to
the
$(0^-,1^-)^{\prime}$ $D$ meson multiplet. We will discuss this feature
in some more detail later. Another notable feature is the
depletion in the differential and total widths of $B\to D\pi\ell\nu$
that arises from the interference between the $(0^-,1^-)$ and
$(0^+,1^+)$ multiplets. In the case of the decay $B\to D^*\pi\ell\nu$,
the depletion caused by this interference at values of $S_{D^*\pi}\le
8.5$ GeV$^2$ is more than compensated by the enhancement that occurs
throughout the rest of phase space. This latter enhancement should
however be taken with caution, as it lies beyond the range of
applicability of our approximation.

\begin{figure}
\let\picnaturalsize=N
\def\picsize{2in}
\def\picfilenamea{fig7a.ps}
\def\picfilenameb{fig7b.ps}
\ifx\nopictures Y\else{\ifx\epsfloaded Y\else\input epsf \fi
\let\epsfloaded=Y
\centerline{
\ifx\picnaturalsize N\epsfxsize \picsize\fi \epsfbox{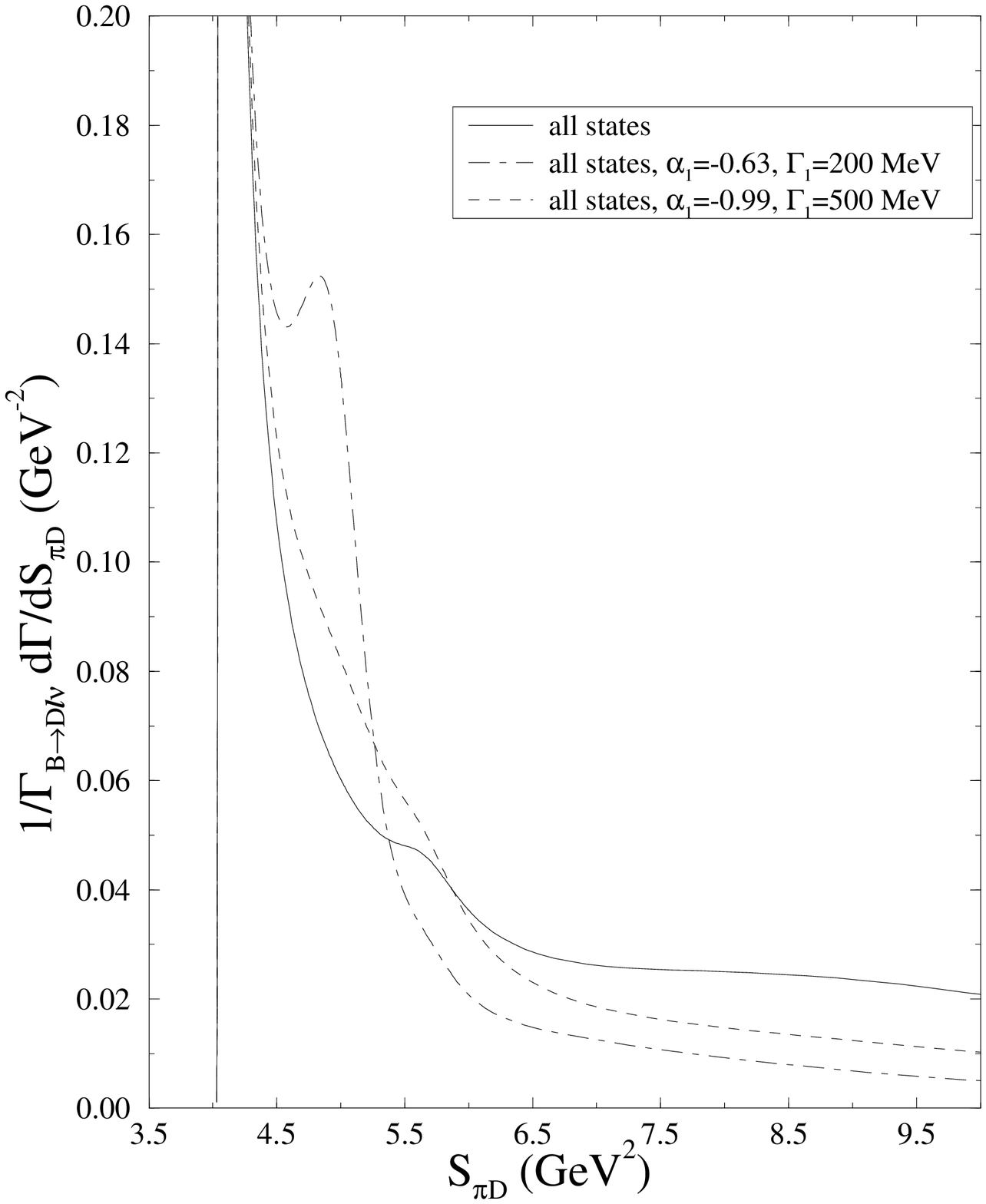}
\hfil
\ifx\picnaturalsize N\epsfxsize \picsize\fi \epsfbox{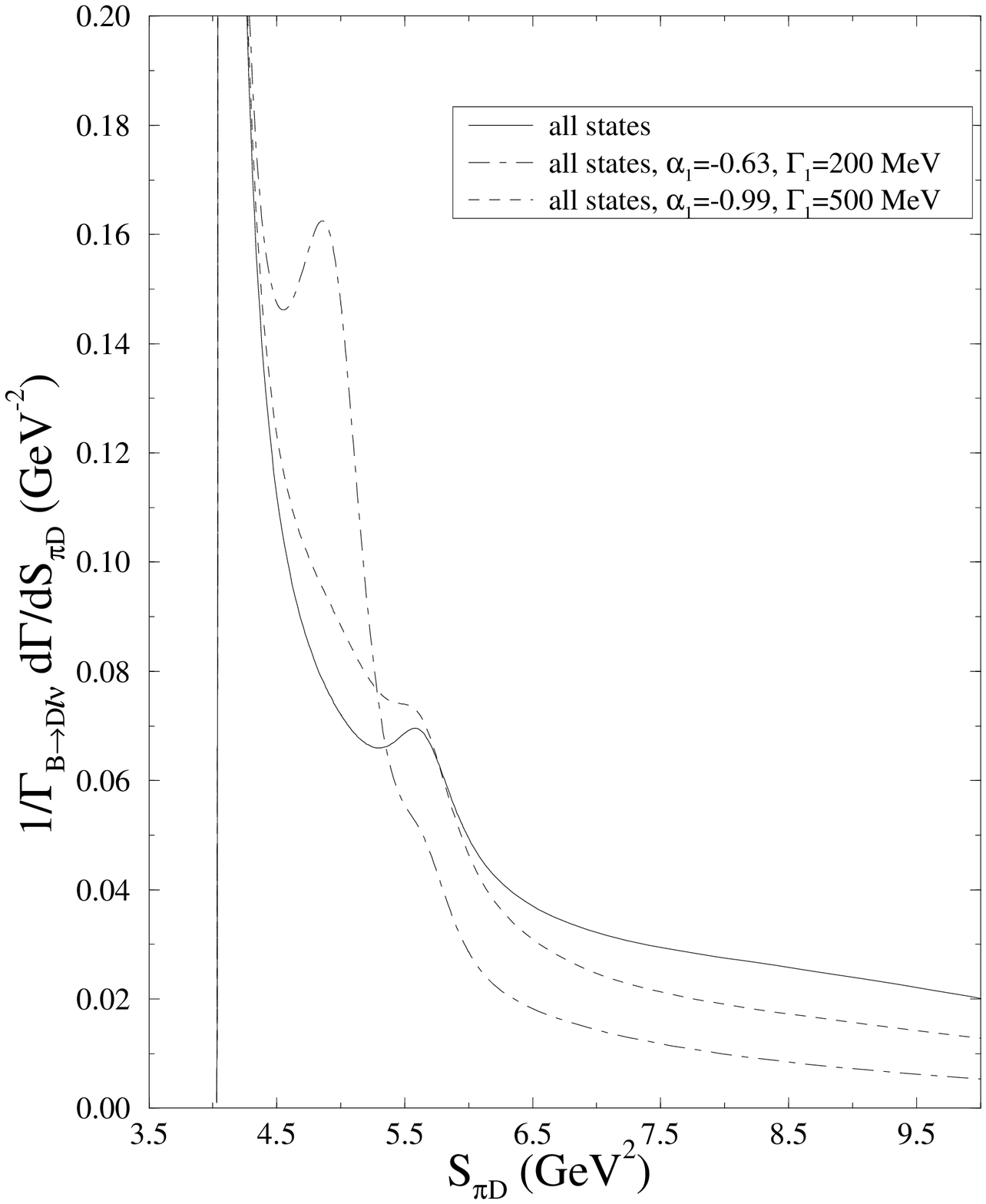}
}}\fi
\caption{The effect on the decay $B\to D \pi^{\pm}  e\nu$,  of
changing the
couplings and total widths of the states in the
$(0^+,1^+)$ multiplet. \label{spectrum3}}
\end{figure}

\begin{figure}
\let\picnaturalsize=N
\def\picsize{2in}
\def\picfilenamea{fig8a.ps}
\def\picfilenameb{fig8b.ps}
\ifx\nopictures Y\else{\ifx\epsfloaded Y\else\input epsf \fi
\let\epsfloaded=Y
\centerline{
\ifx\picnaturalsize N\epsfxsize \picsize\fi \epsfbox{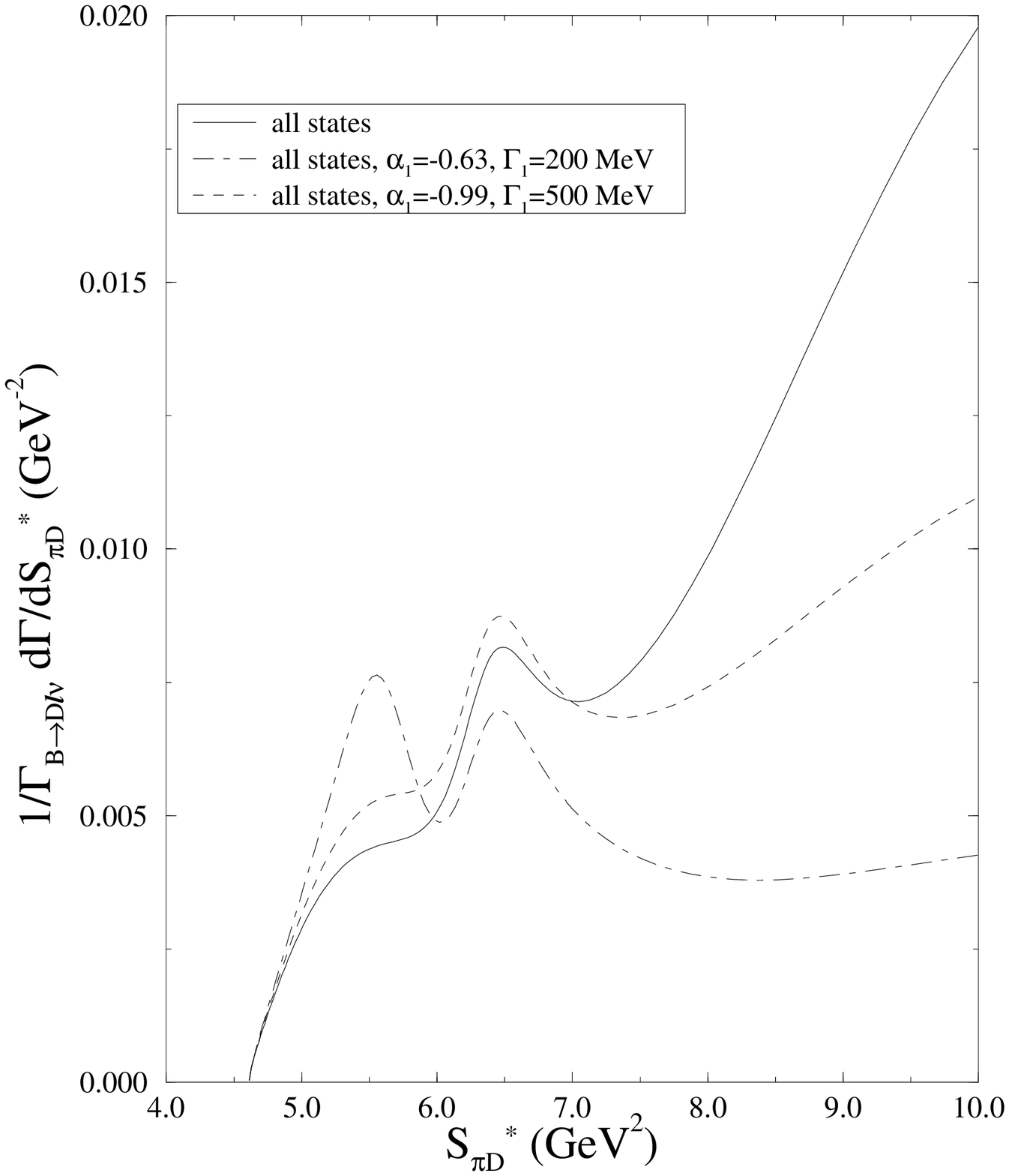}
\hfil
\ifx\picnaturalsize N\epsfxsize \picsize\fi \epsfbox{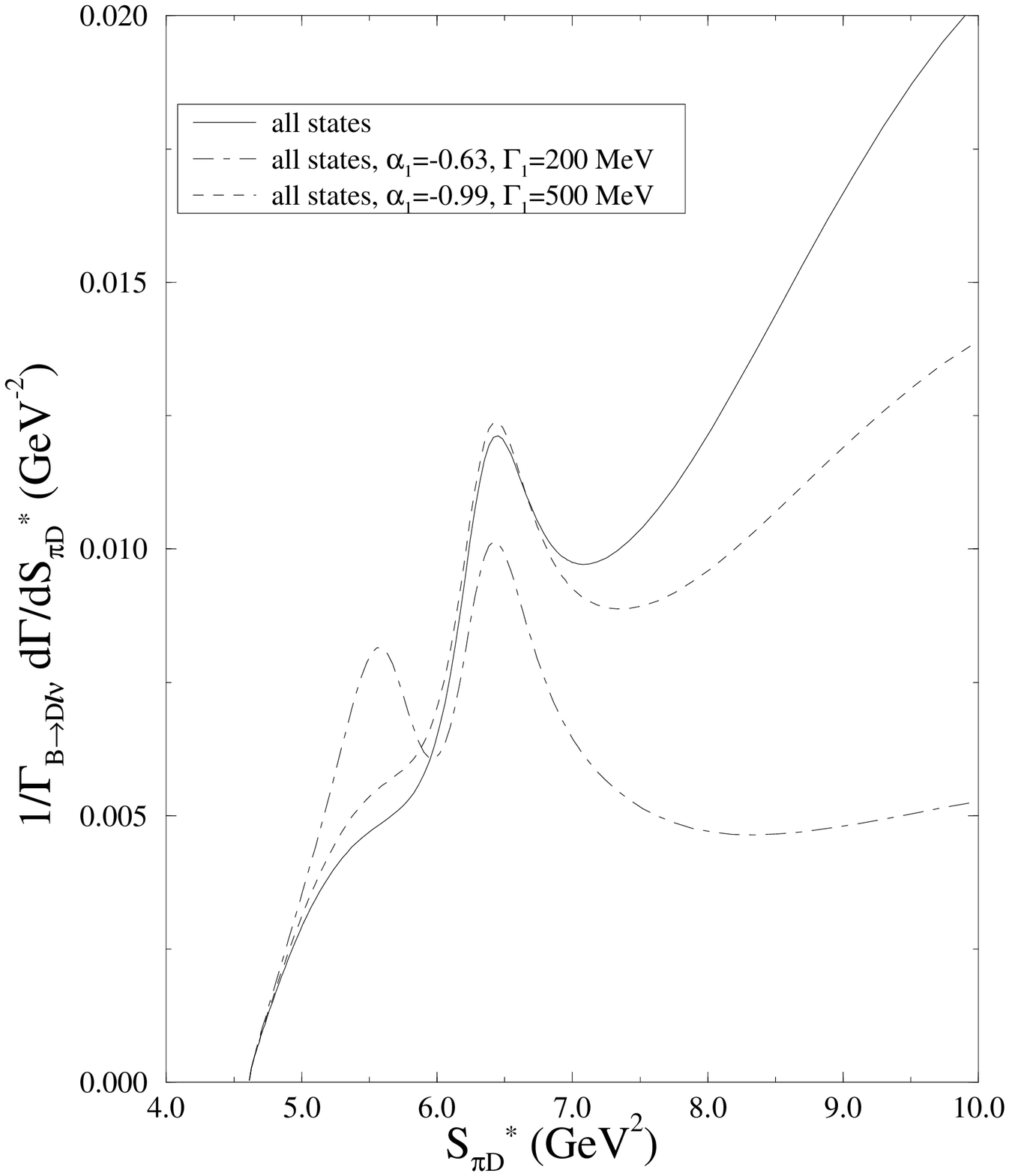}
}}\fi
\caption{The effect on the decay $B\to D^{*}\pi^{\pm}  e\nu$, of
changing the
 couplings and total widths of the states in the
$(0^+,1^+)$ multiplet. \label{spectrum4}}
\end{figure}

\begin{figure}
\let\picnaturalsize=N
\def\picsize{2in}
\def\picfilenamea{fig9a.ps}
\def\picfilenameb{fig9b.ps}
\ifx\nopictures Y\else{\ifx\epsfloaded Y\else\input epsf \fi
\let\epsfloaded=Y
\centerline{
\ifx\picnaturalsize N\epsfxsize \picsize\fi \epsfbox{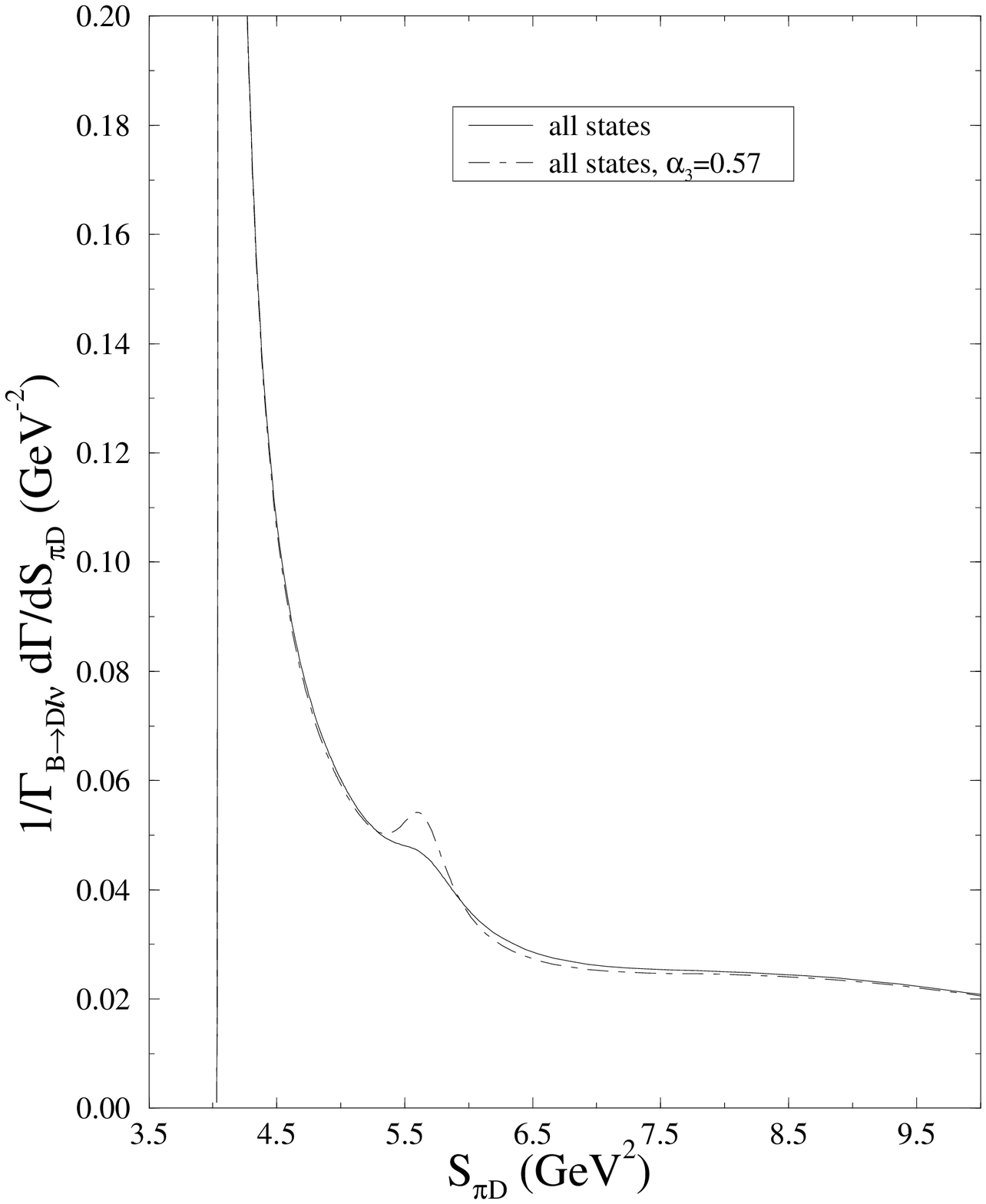}
\hfil
\ifx\picnaturalsize N\epsfxsize \picsize\fi \epsfbox{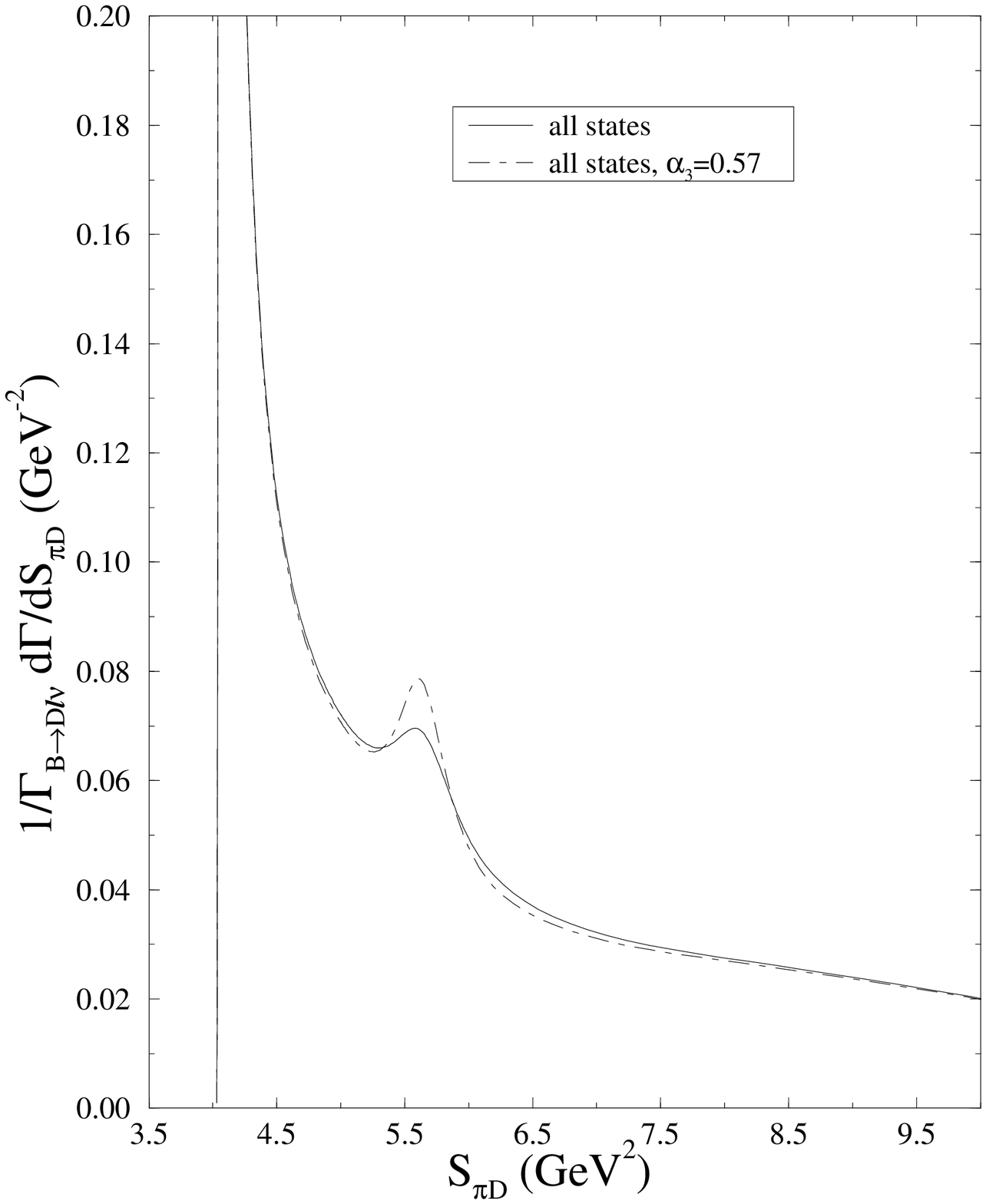}
}}\fi
\caption{The effect on the decay $B\to D\pi^{\pm}  e\nu$, of changing
the
 couplings and total widths of the states in the
$(0^-,1^-)^\prime$ multiplet. \label{spectrum5}}
\end{figure}

\begin{figure}
\let\picnaturalsize=N
\def\picsize{2in}
\def\picfilenamea{fig10a.ps}
\def\picfilenameb{fig10b.ps}
\ifx\nopictures Y\else{\ifx\epsfloaded Y\else\input epsf \fi
\let\epsfloaded=Y
\centerline{
\ifx\picnaturalsize N\epsfxsize \picsize\fi \epsfbox{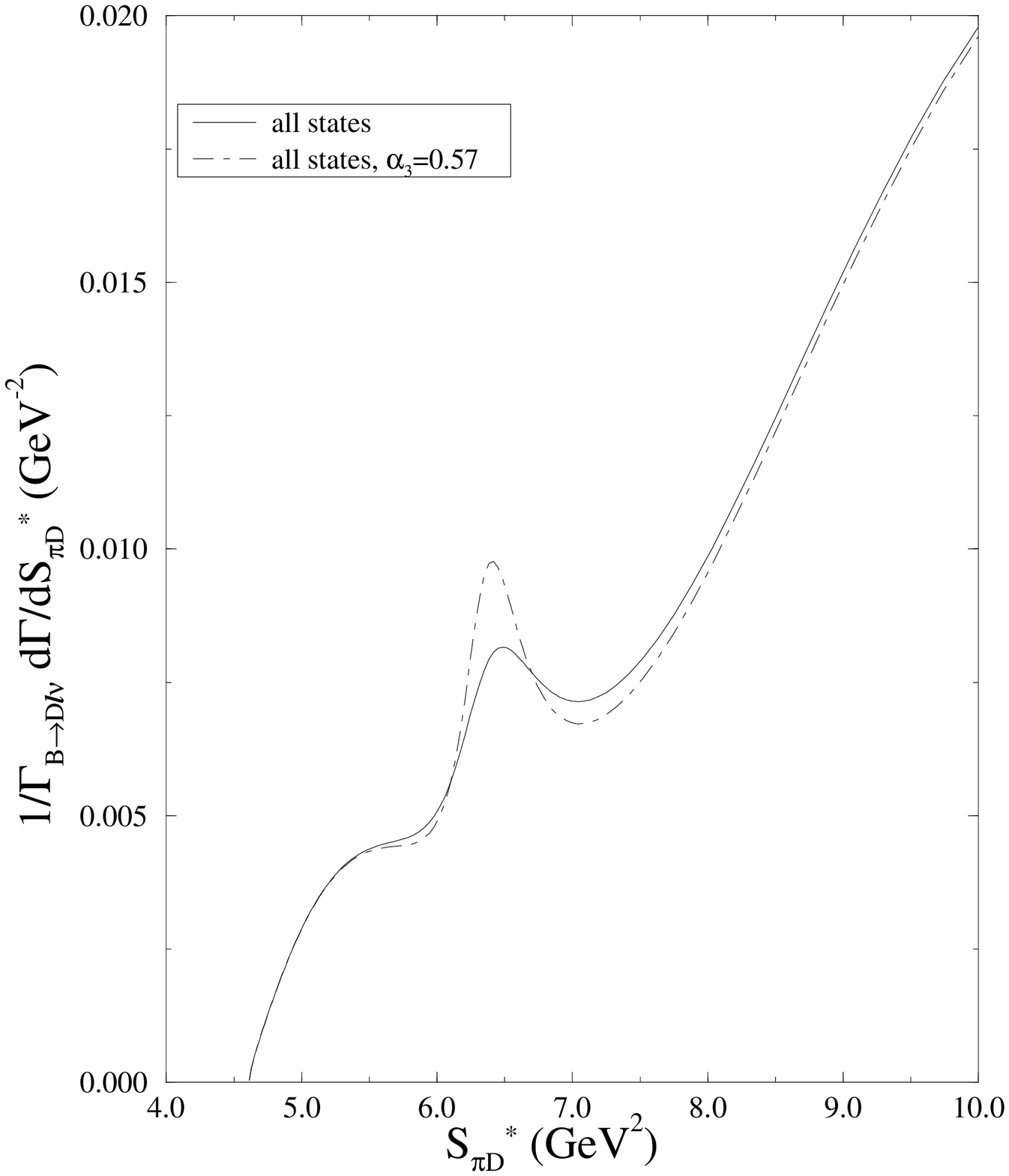}
\hfil
\ifx\picnaturalsize N\epsfxsize \picsize\fi \epsfbox{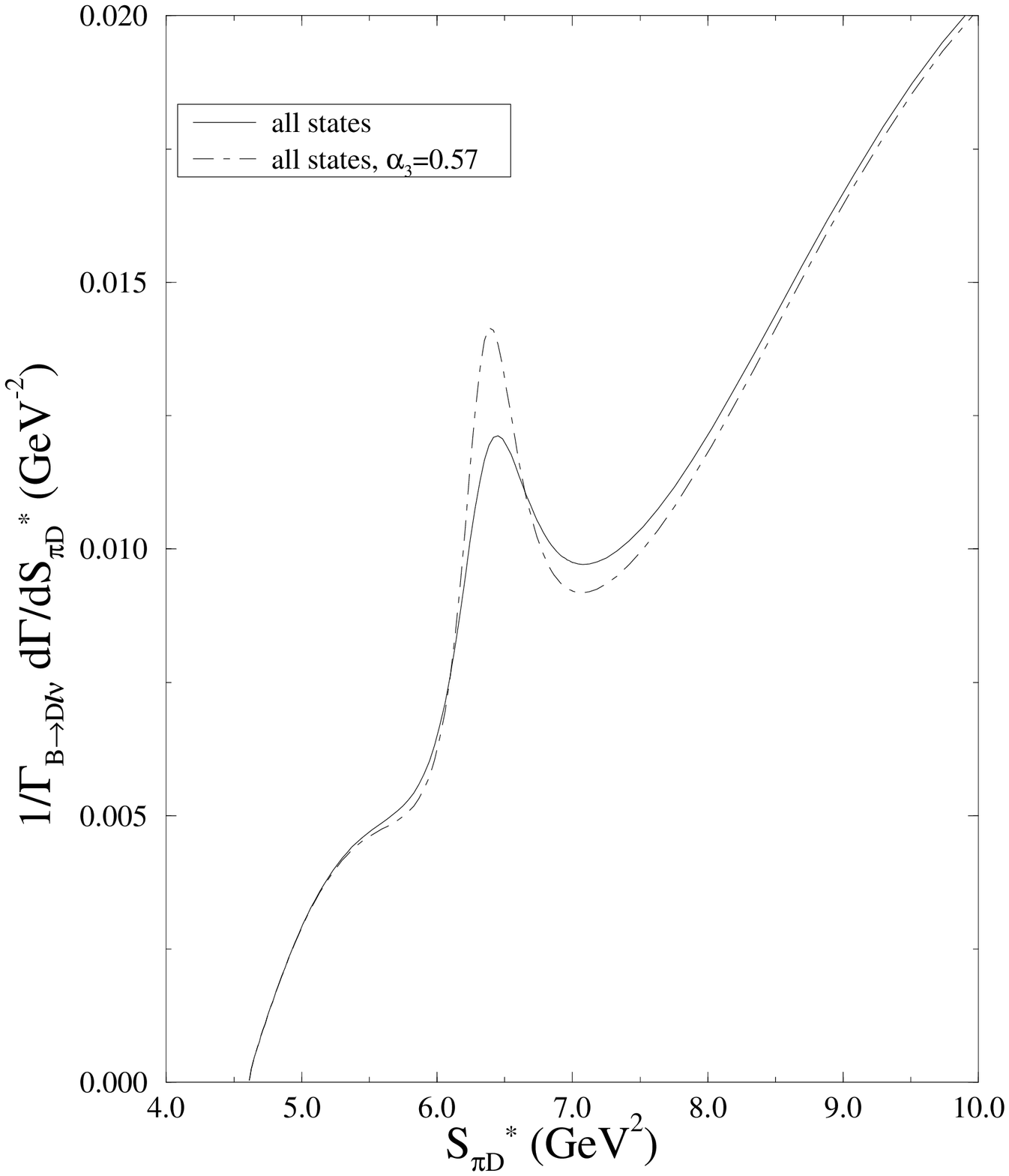}
}}\fi
\caption{The effect on the decay $B\to D^{*}\pi^{\pm}  e\nu$, of
changing the
 couplings and total widths of the states in the
$(0^-,1^-)^\prime$ multiplet. \label{spectrum6}}
\end{figure}

In figures \ref{spectrum3} to \ref{spectrum6} we examine the effects
of
the values of some of the parameters on the spectra. Figures
\ref{spectrum3} and \ref{spectrum4} show the effect of changing the
coupling constant and width of the states in the $(0^+,1^+)$
multiplet.
The values we have obtained in our model are $\alpha_1=-1.43 ~(-1.22)$
and $\Gamma=1.04 ~(0.756)$ GeV. This
width may seem very large, but it is at least consistent with other
model calculations. Nevertheless, we have investigated the effect of
using smaller
widths, namely 500 MeV and 200 MeV. $\alpha_1$ is changed to -0.99 and
-0.63, respectively, in keeping with these changes. We see that the
effect on the
spectrum of $B\to D\pi\ell\nu$
is quite striking, especially at the narrower of the two widths. The
effect on the spectrum of $B\to D^*\pi\ell\nu$ is similarly
striking. The total width of the $B\to D^*\pi\ell\nu$ decay is
strongly
affected by these changes, changing from $1.7 ~(1.7) \times 10^{-15}$
GeV to
$1.0 ~(1.2) \times 10^{-15}$ GeV and $4.9 ~(5.9) \times 10^{-16}$ GeV
as the total width of this multiplet changes from 1.04 ~(0.756) GeV to
500 MeV to 200 MeV. In comparison,
the total width of $B\to D\pi\ell\nu$ is essentially unaffected by
these changes.

In figures \ref{spectrum5} and \ref{spectrum6} we illustrate the
effect
of changing the total width of the $(0^-,1^-)^\prime$ multiplet from
191 ~(174) MeV to 130 MeV, accompanied by a change in
$\alpha_3$ from 0.69 ~(0.66) to 0.57. We see that the narrower state
would provide a clearer signal for experimentalists. The possibility
of
observing this pair of states in $B_{\ell 4}$ decays will depend
strongly on the value of the total width and on $\alpha_3$, as well as
on the effects that various experimental cuts will have on the spectra
we illustrate. While we do not study the effects of cuts in this work,
we can estimate the number of events that one may see in the proposed
$B$-factory.

The integrated width under the peak (from about 5.33 to 6.0 GeV$^2$)
is
$2.7 ~(3.9) \times 10^{-16}$ GeV, corresponding to a branching
fraction
of $5.2 ~(7.4) \times 10^{-4}$, or about $5.2 ~(7.4) \times 10^4$
events, assuming production of $10^{8}$ $B$ mesons. Subtracting the
width that corresponds to a `smooth' background leaves $220 ~(7000)$
events in the peak alone. A similar exercise in the case of the
$D^*\pi\ell\nu$ spectrum yields a width of $7.4 ~(10.3) \times
10^{-17}$ GeV, corresponding to 14200 (19900) events. Removing the
smooth background leaves 1800 (4300) events in the resonant peak.

While the dependence on the model parameters is clear, these estimates
suggest that a study of the spectra of $B\to D\pi\ell\nu$ and $B\to
D^*\pi\ell\nu$ offer some opportunity for discovery or confirmation of
the resonances of the $(0^-,1^-)^{\prime}$ multiplet. Note that if we
include the expected small mass difference between these two states,
the single peak in these figures will become two peaks that are very
close together (separated by about 0.32 GeV$^2$ if the mass splitting
is about 60 MeV). The net effect would be a broadening of the
structure
that we have in our spectra.

In conclusion, we have studied $B_{\ell 4}$ decays in the soft-pion
limit using chiral perturbation theory and heavy quark symmetry. The
resonances which  give  leading contributions in this limit have been
included, and shown to be important in determining both the rate and
the shape of the spectrum. The narrow $(0^{-}, 1^{-})^\prime$
resonances show up as a peak in the $S_{D\pi}$ spectrum, and will
likely be difficult to observe in $B\to D\pi\ell\nu$. The possibility
for detection or confirmation in $B\to D^*\pi\ell\nu$ is more
promising. The wider resonances of the $(0^{+},1^{+})$ multiplet show
some effect on the total rate, but are not likely to be identified
from
the spectrum. Preliminary indications are that they may be identified
at Aleph using the topology of the $B_{\ell 4}$ decays \cite{scott}.
The
effect of these broad resonances is much more pronounced for $B\to
D^*\pi\ell\nu$ than for $B\to D\pi\ell\nu$, although the effects on
both spectra are quite clear.

\acknowledgements

We thank Nathan Isgur for useful discussions. We also thank I. Scott
and J. Bellantoni for discussing some experimental details. W. R.
acknowledges the hospitality and support of Institut des Sciences
Nucl\`eaires and Universit\`e Joseph Fourier, Grenoble, France, as
well
as that of the CERN theory group.

\newpage

\newpage
\appendix
\section{Strong interaction transition vertices with a single pion}

\begin{figure}
\let\picnaturalsize=N
\def\picsize{5.5in}
\def\picfilenamea{Bl4_fig.ps}
\ifx\nopictures Y\else{\ifx\epsfloaded Y\else\input epsf \fi
\let\epsfloaded=Y
\centerline{
\ifx\picnaturalsize N\epsfxsize \picsize\fi \epsfbox{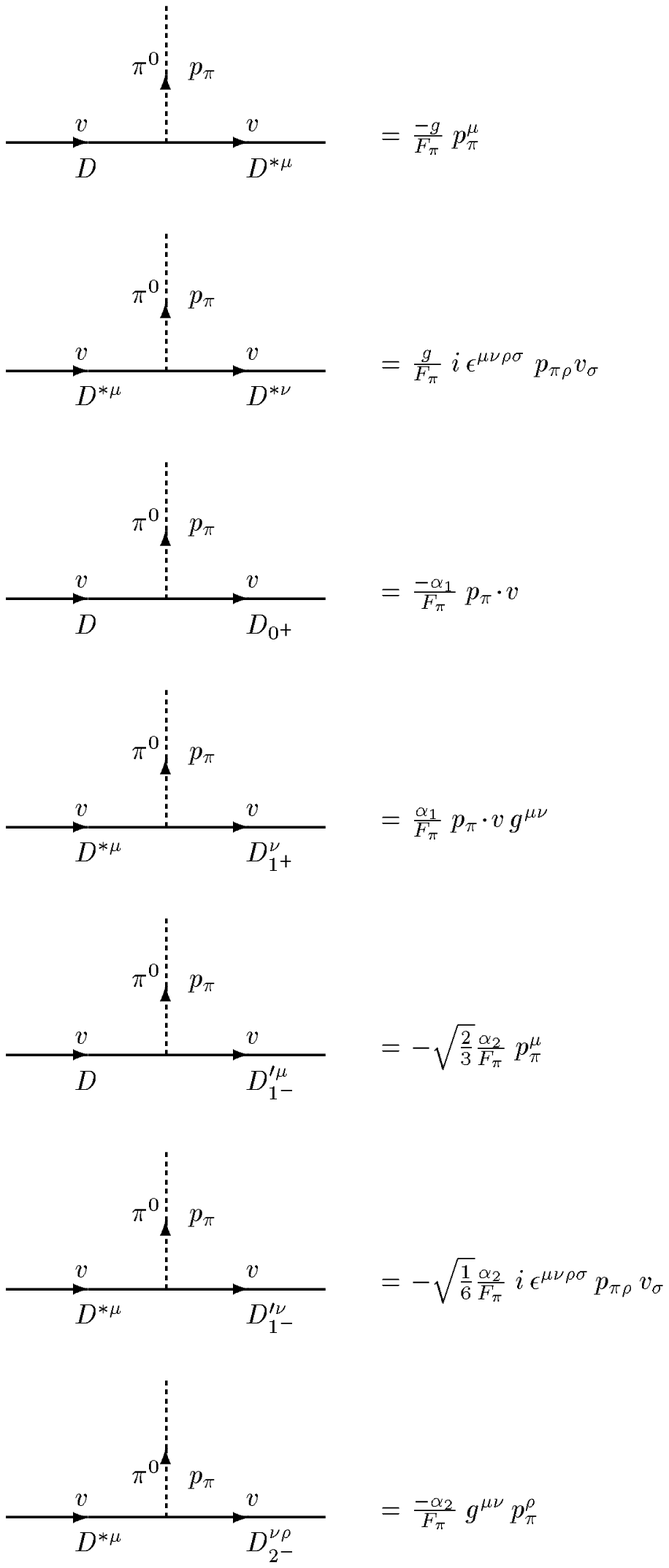}}}
\fi
\vspace*{-1.5cm}
\caption[]{Vertices obtained from the chiral Lagrangian for
soft neutral pion emission. Vertices  for charged pion emission are a
factor
$\sqrt{2}$ of those shown in this figure. \label{vertices}}
\end{figure}

In this appendix, we give  the explicit expressions for the vertices
where one pion emission takes place according to ${\cal L}_{\chi}^{
\rm int}$.
Similar results hold for B mesons.  The vertices are shown in
figure \ref{vertices}.

\section{The charged currents}

In this appendix the explicit expressions for the charged currents
displayed
in eq. (\ref{currents}) are presented.  They are
\begin{eqnarray}
J_{\mu}(0^{-}\rightarrow 0^{-})&=&-\xi(\nu)\, D^{\dagger}(v^{
\prime})(v+v^{\prime})_{\mu}B(v)\nonumber,\\
J_{\mu}(0^{-}\rightarrow 1^{-})&=&\xi(\nu)\,D^{\ast\dagger\nu}(v^{
\prime})\left[i\epsilon_{\nu\mu\alpha\beta}v^{\prime\alpha}v^{
\beta}+g_{\mu\nu}(1+\nu)
-v^{\prime}_{\mu}v_{\nu}\right]B(v),\nonumber\\
J_{\mu}(1^{-}\rightarrow 0^{-})&=&\xi(\nu)\,D^\dagger(v^{\prime})
\left[i\epsilon_{\mu\alpha\beta\rho}v^{\alpha}v^{\prime\beta}+g_{\mu
\rho}(1+\nu)-v_{\mu}
v^{\prime}_{\rho}\right]B^{\ast\rho}(v),\nonumber\\
J_{\mu}(1^{-}\rightarrow 1^{-})&=&\xi(\nu)\,D^{\ast\dagger\nu}(v^{
\prime})\nonumber\\
&\times&\left[g_{\nu\rho}(v+v^{\prime})_{\mu}-g_{\mu\rho}v_{\nu}-g_{
\mu\nu}v^{\prime}_{\rho}+i\epsilon_{\mu\alpha\rho\nu}
(v+v^{\prime})^{\alpha}\right]B^{\ast\rho}(v),\nonumber\\
J_{\mu}(0^{-}\rightarrow 0^{+})&=& -\rho_{1} (\nu)\; D_{+}^{
\dagger}(v^{\prime})(v-v^{\prime})_{\mu}B(v),\nonumber\\
J_{\mu}(0^{-}\rightarrow 1^{+})&=&\rho_{1}(\nu)\;D_{+}^{\ast\dagger
\nu}(v^{\prime})\left[i\epsilon_{\mu\nu\alpha\beta}v^{\alpha}v^{
\prime\beta}
+g_{\mu\nu}(\nu-1)-v^{\prime}_{\mu}v_{\nu}\right]B(v),\nonumber\\
J_{\mu}(1^{-}\rightarrow 0^{+})&=&\rho_{1}(\nu)\;D_{+}^{\dagger}(v^{
\prime})\left[-i\epsilon_{\mu\alpha\beta\rho}v^{\alpha}v^{\prime
\beta}
-g_{\mu\rho}(\nu-1)+v_{\mu}v^{\prime}_{\rho}\right]B^{\ast\rho}(v),
\nonumber\\
J_{\mu}(1^{-}\rightarrow 1^{+})&=&\rho_{1}(\nu)\;D_{+}^{\ast\dagger
\nu}(v^{\prime})\nonumber\\
&\times&\left[-g_{\nu\rho}(v-v^{\prime})_{\mu}+g_{\mu\rho}v_{\nu}
-g_{\mu\nu}v^{\prime}_{\rho}+i\epsilon_{\nu\mu\rho\alpha}(-v+v^{
\prime})^{\alpha}\right]B^{\ast\rho}(v),\nonumber\\
J_{\mu}(0^{-}\rightarrow 1^{-})&=&\frac{1}{\sqrt{6}}\rho_{2}(\nu)
\,D_{1^{-}}^{\dagger\nu}(v^{\prime})\nonumber\\
&\times&\left[i\epsilon_{\mu\nu\alpha\beta}v^{\alpha}v^{\prime\beta}
(\nu-1)+g_{\mu\nu}(\nu^{2}-1)-v_{\nu}((2+\nu)v^{\prime}_{\mu}-3v_{
\nu})\right]B(v),\nonumber\\
J_{\mu}(0^{-}\rightarrow 2^{-})&=&-\rho_{2}(\nu)\,D_{2^{-}}^{\ast
\dagger\nu\rho}(v^{\prime})\nonumber\\
&\times&\left[- v_{\nu} g_{\mu\rho}(\nu-1)
+v_{\nu}v_{\rho}v^{\prime}_{\mu}+ i\epsilon_{\mu\nu\alpha\beta} v_{
\rho}v^{\alpha}v^{\prime\beta}\right]B^{\ast\sigma}(v),\nonumber\\
J_{\mu}(1^{-}\rightarrow 1^{-})&=&\frac{1}{\sqrt{6}}\rho_{2}(\nu)
\,D_{1^{-}}^{\dagger\nu}(v^{\prime})\nonumber\\
&\times&\left[(v+v^{\prime})_{\mu}g_{\nu\sigma}(\nu-1)-3v_{\nu}v_{
\mu}v^{\prime}_{\sigma}+
2v_{\nu}g_{\mu\sigma}(\nu-1)-g_{\mu\nu}v^{\prime}_{\sigma}(\nu-1)
\right.\nonumber\\
&-&\left.i\epsilon_{\mu\nu\alpha\sigma}(v-v^{\prime})^{\alpha}(1+\nu)+
2i\epsilon_{\nu\sigma\alpha\beta}v^{\alpha}v_{\mu}v^{\prime\beta}+
i\epsilon_{\mu\sigma\alpha\beta}v_{\nu}v^{\alpha}v^{\prime\beta}
\right]B^{\ast \sigma}(v),\nonumber\\
J_{\mu}(1^{-}\rightarrow 2^{-})&=&\rho_{2}(\nu)\,D_{2^{-}}^{\ast
\dagger\nu\rho}(v^{\prime})\nonumber\\
&\times&\left[-i \epsilon_{\mu\nu\delta\sigma}
v_{\rho}(v-v^{\prime})^{\delta}+g_{\mu\sigma}v_{\nu}v_{\rho}-g_{\mu
\rho}v_{\nu}v^{\prime}_{\sigma}-g_{\rho\sigma}v_{\nu}(v-v^{\prime})_{
\mu}\right]
B^{\ast \sigma}(v),\nonumber\\
J_{\mu}(0^{-}\rightarrow 0^{-\prime})&=&-\xi^{(1)}(\nu)\, D^{\prime
\dagger}(v^{\prime})(v+v^{\prime})_{\mu}B(v)\nonumber,\\
J_{\mu}(0^{-}\rightarrow 1^{-\prime})&=&\xi^{(1)}(\nu)\,D^{\prime\ast
\dagger\nu}(v^{\prime})\left[i\epsilon_{\nu\mu\alpha\beta}v^{\prime
\alpha}v^{\beta}+g_{\mu\nu}(1+\nu)
-v^{\prime}_{\mu}v_{\nu}\right]B(v),\nonumber\\
J_{\mu}(1^{-}\rightarrow 0^{-\prime})&=&\xi^{(1)}(\nu)\,D^{\prime
\dagger}(v^{\prime})\left[i\epsilon_{\mu\alpha\beta\rho}v^{\alpha}v^{
\prime\beta}+g_{\mu\rho}(1+\nu)-v_{\mu}
v^{\prime}_{\rho}\right]B^{\ast\rho}(v),\nonumber\\
J_{\mu}(1^{-}\rightarrow 1^{-\prime})&=&\xi^{(1)}(\nu)\,D^{\prime\ast
\dagger\nu}(v^{\prime})\nonumber\\
&\times&\left[g_{\nu\rho}(v+v^{\prime})_{\mu}-g_{\mu\rho}v_{\nu}-g_{
\mu\nu}v^{\prime}_{\rho}+i\epsilon_{\mu\alpha\rho\nu}
(v+v^{\prime})^{\alpha}\right]B^{\ast\rho}(v).
\end{eqnarray}

The effective currents where a $B$-meson resonance decays into a
ground state  $D$-meson
are easily obtained  simply taking the
hermitian (minus the hermitian) conjugate of the vector (axial-vector)
portion of the currents displayed above followed by the interchange
of symbols $B\leftrightarrow D$ and $v\leftrightarrow v^{\prime}$.

\section{The Form Factors}
In this appendix we give the form factors needed in eqns. (
\ref{omega1}) and (\ref{omegamunu1}).
The non-resonant and the resonant contributions are displayed
separately.

\subsection{$B\rightarrow D \pi \ell  \bar{\nu} $}

If we write
\begin{equation}
\Omega_{\mu}=- i h\;M_{B}M_{D}\,\epsilon_{\mu\nu\rho\sigma}
v^{\nu}v^{\prime\rho} p_{\pi}^{\sigma}+A_{1}\; p_{\pi\mu}+
A_{2} \;M_{B} v_{\mu}+A_{3}\; M_{D} v^{\prime}_{\mu},
\end{equation}
the form factors in equation (\ref{omega1}) are
\begin{eqnarray}
H&=&-h,  \nonumber\\
F&=&A_{2}+\frac{1}{2} (A_{1}+A_{3}),\nonumber\\
G&=&\frac{1}{2} (A_{3}-A_{1}),\nonumber\\
R&=& A_{2}.
\end{eqnarray}
{}From eqn. (\ref{nonresonant1}) we obtain the non-resonant
contributions to the form factors as
\begin{eqnarray}\label{nonresonant3}
h_{NR}&=&    \frac{g}{2 F_{0}}\; \frac{\xi(\nu)}{M_{B}M_{D}}\;
\left( \frac{1}{p_{\pi}     \!\cdot\! v+\delta m_{B}-i\epsilon}-
\frac{1}{p_{\pi}     \!\cdot\! v^{\prime}-\delta m_{D}+i\epsilon}
\right),  \nonumber\\
A_{ 1\, NR}&=& - \frac{g}{2 F_{0}}\; \xi(\nu)\;(1+  \nu)
\left( \frac{1}{p_{\pi}     \!\cdot\! v +\delta m_{B}-i\epsilon}-
\frac{1}{p_{\pi}     \!\cdot\! v^{\prime}-\delta m_{D}+i\epsilon}
\right),\nonumber \\
A_{2\,  NR}&=& \frac{g}{2 F_{0}} \;\frac{\xi(\nu)}{M_{B}}\;
\left( \frac{p_{\pi}     \!\cdot\! v+ p_{\pi}     \!\cdot\! v^{
\prime}}{p_{\pi}     \!\cdot\! v+\delta m_{B}-i\epsilon }\right),
\nonumber\\
A_{3\,  NR}&=& -\frac{g}{2 F_{0}}\; \frac{\xi(\nu)}{M_{D}}
\left(   \frac{p_{\pi}     \!\cdot\! v+ p_{\pi}     \!\cdot\! v^{
\prime}}{p_{\pi}     \!\cdot\! v^{\prime}-\delta m_{D}+i\epsilon }
\right).
\end{eqnarray}
These results are the same as those obtained by Lee and collaborators
\cite{wiselee}, and by Cheng and collaborators \cite{chengetal}.
\begin{figure}
\let\picnaturalsize=N
\def\picsize{5.5in}
\def\picfilenamea{Bl4fig11.ps}
\ifx\nopictures Y\else{\ifx\epsfloaded Y\else\input epsf \fi
\let\epsfloaded=Y
\centerline{
\ifx\picnaturalsize N\epsfxsize \picsize\fi \epsfbox{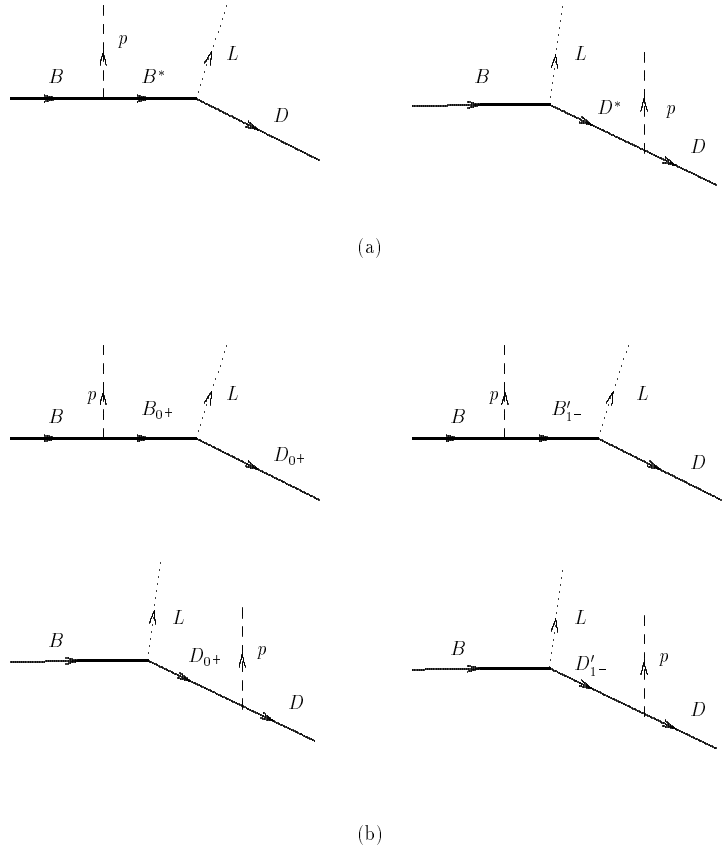}}}
\fi
\vspace*{-2.5cm}
\caption{Non-resonant (a) and resonant (b) diagrams contributing
to $B\rightarrow D \pi \ell \bar{\nu}$.
The dashed line represents the pion and the dotted line which
emerges from the electroweak charged current
carries the momentum of the $\ell \bar{\nu}$ pair.
 Horizontal  solid lines correspond to
four velocity $v$  and the oblique  ones
to $v^{\prime}$. \label{bdpienu}}
\end{figure}

{}From eqn. (\ref{resonant1}) the resonant contributions are
\begin{eqnarray}\label{resonant3}
h_{R}&=& \frac{\alpha_{2} \; \rho_{2}( \nu)}{6 F_{0} M_{B}M_{D}}\;
\,( \nu-1) \left( \frac{1}{p_{\pi}     \!\cdot\! v+ \delta
\tilde{m}_{2} }-
 \frac{1}{p_{\pi}     \!\cdot\! v^{\prime}- \delta \tilde{m}_{2} }
\right) \nonumber \\
&+& \frac{\alpha_{3}}{2 F_{0}}\; \frac{\xi^{(1)}(\nu)}{M_{B}M_{D}}\;
\left( \frac{1}{p_{\pi}     \!\cdot\! v+ \delta \tilde{m}_{3} }-
\frac{1}{p_{\pi}     \!\cdot\! v^{\prime}- \delta \tilde{m}_{3} }
\right),\nonumber\\
A_{ 1\, R}&=& -\frac{\alpha_{2}\;  \rho_{2}( \nu)}{6 F_{0}}\;
(  \nu^{ 2}-1)\; \left( \frac{1}{p_{\pi}     \!\cdot\! v+ \delta
\tilde{m}_{2} }-
 \frac{1}{p_{\pi}     \!\cdot\! v^{\prime}- \delta \tilde{m}_{2} }
\right)\nonumber\\
&-& \frac{\alpha_{3}\,\xi^{(1)}(\nu)}{2 F_{0}} \;(1+  \nu)
\left( \frac{1}{p_{\pi}     \!\cdot\! v+ \delta \tilde{m}_{3} }-
\frac{1}{p_{\pi}     \!\cdot\! v^{\prime}- \delta \tilde{m}_{3} }
\right), \nonumber\\
A_{2\, R}&=&  \frac{\alpha_{1}\;  \rho_{1}
( \nu)}{2F_{0} M_{B}} \left( \frac{p_{\pi}     \!\cdot\! v^{\prime}}
{p_{\pi}     \!\cdot\! v^{\prime}- \delta \tilde{m}_{1} }+\frac{p_{
\pi}     \!\cdot\! v}
{p_{\pi}     \!\cdot\! v+ \delta \tilde{m}_{1} }\right)\nonumber \\
&+&
\frac{\alpha_{2}\;  \rho_{2}( \nu)}{ F_{0} M_{B}}\;
\left\{\frac{1}{p_{\pi}     \!\cdot\! v+ \delta \tilde{m}_{2} }
\left[\frac{1}{6} (\nu\; p_{\pi}     \!\cdot\! v^{\prime} -p_{\pi}
 \!\cdot\! v)+
\frac{1}{3}(p_{\pi}     \!\cdot\! v^{\prime}- \nu\;p_{\pi}     \!
\cdot\! v  )\right]\right.\nonumber\\
&+&\left.\frac{1}{2} \frac{1}{p_{\pi}     \!\cdot\! v^{\prime}-
\delta \tilde{m}_{2} }
(p_{\pi}     \!\cdot\! v-\nu\;p_{\pi}     \!\cdot\! v^{\prime} )
\right\}
\nonumber\\
&+&\frac{\alpha_{3}}{2 F_{0}}\; \frac{\xi^{(1)}(\nu)}{M_{B}}\;
\left( \frac{p_{\pi}     \!\cdot\! ( v+v^{\prime})}{p_{\pi}     \!
\cdot\! v+ \delta \tilde{m}_{3} }\right),\nonumber\\
A_{3\, R}&=& -\frac{\alpha_{1}\;  \rho_{1}
( \nu)}{2F_{0} M_{D}} \left( \frac{p_{\pi}     \!\cdot\! v^{\prime}}
{p_{\pi}     \!\cdot\! v^{\prime}- \delta \tilde{m}_{1} }+\frac{p_{
\pi}     \!\cdot\! v}
{p_{\pi}     \!\cdot\! v+ \delta \tilde{m}_{1} }\right)\nonumber\\
&-&
\frac{\alpha_{2} \; \rho_{2}( \nu)}{ F_{0} M_{D}}\;
\left\{\frac{1}{p_{\pi}     \!\cdot\! v^{\prime}- \delta
\tilde{m}_{2} }
\left[\frac{1}{6} (\nu\;p_{\pi}     \!\cdot\! v -p_{\pi}     \!\cdot
\! v^{\prime})+
\frac{1}{3}(p_{\pi}     \!\cdot\! v-\nu\;p_{\pi}     \!\cdot\! v^{
\prime} )\right]\right.\nonumber\\
&+&\left.\frac{1}{2}\; \frac{1}{p_{\pi}     \!\cdot\!  v+ \delta
\tilde{m}_{2} }
(p_{\pi}     \!\cdot\! v^{\prime} -\nu\;p_{\pi}     \!\cdot\! v )
\right\}
\nonumber\\
&-&\frac{\alpha_{3}}{2 F_{0}}\; \frac{\xi^{(1)}(\nu)}{M_{D}}
\;\left(  \frac{p_{\pi}     \!\cdot\! (v+v^{\prime})}{p_{\pi}     \!
\cdot\! v^{\prime}- \delta \tilde{m}_{3} }\right) .
\end{eqnarray}

\subsection{$B\rightarrow D^{\ast} \pi \ell \nu $}

The most general form for the tensor $\Omega_{\mu\nu}$ in terms
of the vectors $v_{\mu}$, $v_{\mu}^{\prime}$ and $p_{\pi\mu}$ is
(terms which vanish upon contraction with $\epsilon^{\ast
\nu}_{D}(v^{\prime})$
are not displayed)
\begin{eqnarray}\label{omegamunu2}
\Omega_{\mu\nu}(v,v^{\prime},p_{\pi})&=& \frac{i}{2} \;\epsilon_{\mu
\nu\rho\sigma}\;
\left[ h_{1}\; M_{B} M_{D} v^{\rho} v^{\prime\sigma}+
h_{2}\; M_{B} v^{\rho}p_{\pi}^{\sigma}+
h_{3}\; M_{D} v^{\prime \rho}p_{\pi}^{\sigma}\right]\nonumber\\
&+& f_{1}\; M_{B} v_{\mu} p_{\pi\nu}+f_{2}\; M_{D} v^{\prime}_{\mu}
p_{\pi\nu}
+f_{3}\; p_{\pi\mu} p_{\pi\nu}+
f_{4}\; M_{B}^{2} v_{\mu} v_{\nu}\nonumber\\
&+&f_{5}\; M_{B} M_{D} v^{\prime}_{\mu} v_{\nu}+f_{6} \;M_{B} p_{\pi
\mu} v_{\nu}+
k \;g_{\mu\nu} \nonumber\\
&+&\frac{i}{2}\; \epsilon_{\mu\delta\rho\sigma}
\;v^{\delta}v^{\prime \rho} p_{\pi}^{\sigma} \;\left( g_{1}\; p_{\pi
\nu}+g_{2}\; M_{B}  v_{\nu}\right)
\nonumber\\
&+&\frac{i}{2}\; \epsilon_{\nu\delta\rho\sigma} \;v^{\delta}v^{\prime
\rho} p_{\pi}^{\sigma} \;\left( g_{3}\; M_{B}v_{\mu}+g_{4}\; M_{D}
  v^{\prime}_{\mu}+g_{5}\;p_{\pi\mu} \right).
\end{eqnarray}
The form factors appearing in eqn. (\ref{omegamunu1}) are related to
the ones in this expression
by
\begin{eqnarray}
H_{1}&=&      \frac{1}{2}\;( h_{1}-  h_{2}-h_{3}),\nonumber\\
H_{2}&=&  - \frac{1}{2}\;(   h_{1}+ h_{2}),\nonumber\\
H_{3}&=& \frac{1}{2}\;(  - h_{1}+ h_{2}),\nonumber\\
F_{1}&=&\frac{1}{2}\;\left( f_{1}+f_{2}+\frac{1}{2} f_{3}+f_{4}+
\frac{1}{2} f_{5}+
\frac{1}{2} f_{6}\right),\nonumber\\
F_{2}&=&\frac{1}{4}\; \left(f_{2}-f_{3}+f_{5} -f_{6}\right),\nonumber
\\
F_{3}&=&f_{4}+\frac{1}{2}\;(f_{5}+f_{6}),\nonumber \\
F_{4}&=&\frac{1}{2}\;(f_{5}-f_{6}),\nonumber\\
K&=& k,\nonumber\\
G^{A}_{1}&=&-\frac{1}{4 M_{B}M_{D}}\; (g_{1}+g_{2}),\nonumber\\
G^{A}_{2}&=&-\frac{1}{2M_{B}M_{D}}g_{2},\nonumber\\
G^{B}_{1}&=&-\frac{1}{2M_{B}M_{D}}\;\left(g_{3}+\frac{1}{2}
\;(g_{4}+g_{5})\right),\nonumber\\
G^{B}_{2}&=&-\frac{1}{4 M_{B}M_{D}}\;(g_{4}-g_{5}).
\end{eqnarray}

\begin{figure}
\let\picnaturalsize=N
\def\picsize{5in}
\def\picfilenamea{Bl4fig111.ps}
\ifx\nopictures Y\else{\ifx\epsfloaded Y\else\input epsf \fi
\let\epsfloaded=Y
\centerline{
\ifx\picnaturalsize N\epsfxsize \picsize\fi \epsfbox{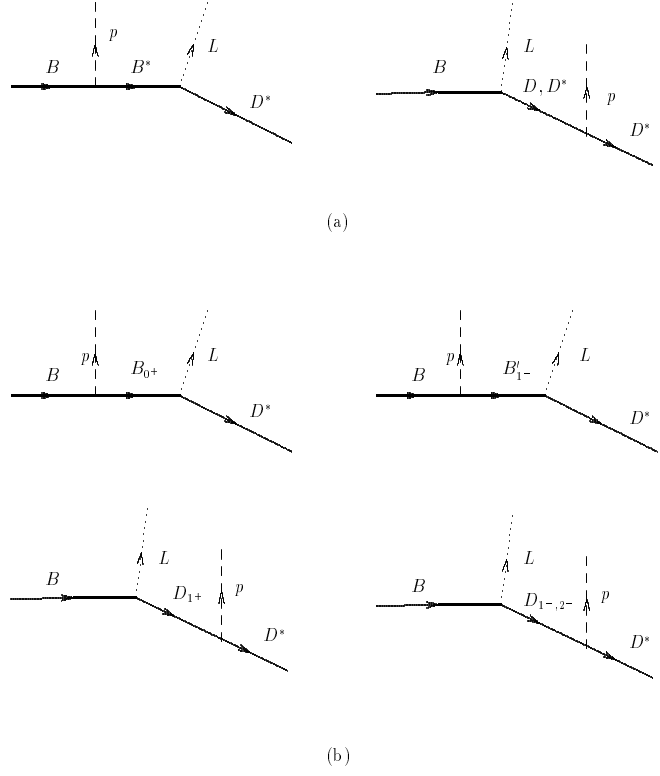}}}
\fi
\vspace*{-2.5cm}
\caption{Non-resonant (a) and resonant (b) diagrams contributing
to $B\rightarrow D^{\ast} \pi \ell \bar{\nu}$. \label{bdstarpienu}}
\end{figure}

{}From eqn. (\ref{nonresonant2}) the non-resonant contributions to the
form factors are
\begin{eqnarray}
h_{1\, NR}&=& -\frac{g \xi(\nu)}{F_{0} M_{B}M_{D}}\;\;\frac{     p_{
\pi}     \!\cdot\! v }{ p_{\pi}     \!\cdot\! v+\delta m_{B}-i
\epsilon}
,\nonumber\\
h_{2\, NR}&=& -\frac{g \xi(\nu)}{F_{0} M_{B}}\;\;\frac{1}{p_{\pi}
\!\cdot\! v+\delta m_{B}-i\epsilon},  \nonumber\\
h_{3\, NR}&=&  -\frac{g \xi(\nu)}{F_{0} M_{D}}\;
\left(\frac{1}{p_{\pi}     \!\cdot\! v +\delta m_{B}-i\epsilon}-
\frac{1+\nu}{p_{\pi}     \!\cdot\! v^{\prime}+i\epsilon }\right),
\nonumber\\
f_{ 1\, NR}&=& -\frac{g \xi(\nu)}
{2 F_{0} M_{B}}\;\left(\frac{1}{p_{\pi}     \!\cdot\! v+\delta
m_{B}-i\epsilon }-\frac{1}
{p_{\pi}     \!\cdot\! v^{\prime}+\delta m_{D}+i\epsilon}\right),
\nonumber\\
f_{2 \, NR}&=& \frac{M_{B}}{M_{D}}\;f_{ 1\, NR},\nonumber\\
f_{ 3\, NR}&=&f_{ 4\, NR}= 0,\nonumber\\
f_{ 5\, NR}&=&\frac{g \xi(\nu)}{2 F_{0} M_{B} M_{D}}\,\left(1+\frac{
 p_{\pi}     \!\cdot\! v }{ p_{\pi}     \!\cdot\! v+\delta m_{B}-i
\epsilon}\right)
, \nonumber\\
f_{6 \, NR}&=&  \frac{g \xi(\nu)}{2 F_{0} M_{B}}\;\left(\frac{1}{p_{
\pi}     \!\cdot\! v +\delta m_{B}-i\epsilon}-
\frac{1}{ p_{\pi}     \!\cdot\! v^{\prime} +i\epsilon  }\right),
\nonumber\\
k_{ NR}&=&  \frac{g     \xi(\nu)}{2\,F_{0}}\;
\left(\frac{p_{\pi}     \!\cdot\! v^{\prime} -\nu\,p_{\pi}     \!
\cdot\! v}{ p_{\pi}     \!\cdot\! v+\delta m_{B}-i\epsilon}
+\frac{p_{\pi}     \!\cdot\! v-\nu\,p_{\pi}     \!\cdot\! v^{
\prime}}{p_{\pi}     \!\cdot\! v^{\prime}+i\epsilon }\right),
\nonumber\\
g_{1 \, NR}&=& 0,\nonumber\\
g_{2 \, NR}&=& 0,\nonumber\\
g_{3 \, NR}&=&0, \nonumber\\
g_{ 4\, NR}&=&\frac{g  \xi(\nu)}{F_{0}\,M_{D}}\;\;\frac{1}{\,p_{\pi}
   \!\cdot\! v^{\prime}+i\epsilon  } ,\nonumber\\
g_{5 \, NR}&=& 0 .
\end{eqnarray}
These results are the same as those obtained by other authors
\cite{wiselee,chengetal}.

Finally, the resonant contributions are obtained from eqn. (
\ref{resonant2}) and are
\begin{eqnarray}
h_{1\, R}&=&  \frac{\alpha_{1} \; \rho_{1} (\nu)}{F_{0} M_{B} M_{D}}\;
\left(\frac{p_{\pi}     \!\cdot\! v}{-p_{\pi}     \!\cdot\! v- \delta
\tilde{m}_{1} }-
\frac{p_{\pi}     \!\cdot\! v^{\prime}}{p_{\pi}     \!\cdot\! v^{
\prime}- \delta \tilde{m}_{1} }\right)\nonumber\\
&+& \frac{\alpha_{2}\;  \rho_{2}(\nu)}
{3F_{0} M_{B} M_{D}} \;
\left(\frac{p_{\pi}     \!\cdot\! v\, (1+2\nu)-p_{\pi}     \!\cdot\!
v^{\prime}}
{p_{\pi}     \!\cdot\! v+ \delta \tilde{m}_{2} }+\frac{3}{2}\frac{\nu
\;p_{\pi}     \!\cdot\! v^{\prime} -
p_{\pi}     \!\cdot\! v}{p_{\pi}     \!\cdot\! v^{\prime}- \delta
\tilde{m}_{2} }\right)\nonumber\\
&-&\frac{\alpha_{3} \;\xi^{(1)}(\nu)}
{F_{0} M_{B}M_{D}} \frac{p_{\pi}     \!\cdot\! v}{p_{\pi}     \!\cdot
\! v+ \delta \tilde{m}_{3} },\nonumber\\
h_{2\, R}&=& \frac{\alpha_{2}\; (1+\nu) \; \rho_{2}(\nu)}
{3 F_{0}M_{B}} \frac{1}{- p_{\pi}     \!\cdot\! v-  \delta
\tilde{m}_{2} }
-\frac{\alpha_{3}\;\xi^{(1)}(\nu)}{F_{0} M_{B}\;(p_{\pi}     \!\cdot
\! v+ \delta \tilde{m}_{3} )}, \nonumber\\
h_{3\, R}&=& \frac{\alpha_{2} \;  \rho_{2}(\nu)}{3 F_{0} M_{D}}\;
\left(\frac{1+\nu}{p_{\pi}     \!\cdot\! v+  \delta \tilde{m}_{2} }
- \frac{\nu^{2}-1}{2 (p_{\pi}     \!\cdot\! v^{\prime}- \delta
\tilde{m}_{2} )}\right)\nonumber\\
& -&\frac{\alpha_{3} \;
\xi^{(1)}(\nu)}{F_{0} M_{D}}\;\left(
\frac{1}{p_{\pi}     \!\cdot\! v+ \delta \tilde{m}_{3} }-\frac{1+\nu}{
p_{\pi}     \!\cdot\! v^{\prime}- \delta \tilde{m}_{3} }\right),
\nonumber\\
f_{ 1\, R}&=& -\frac{ \alpha_{2} \; \rho_{2}(\nu)\, (\nu-1)}
{6 F_{0}M_{B}} \;\left(\frac{1}{p_{\pi}     \!\cdot\! v+  \delta
\tilde{m}_{2} }-
\frac{1}{p_{\pi}     \!\cdot\! v^{\prime}- \delta \tilde{m}_{2} }
\right) \nonumber\\
&-&\frac{\alpha_{3} \;\xi^{(1)}(\nu)}{2 F_{0} M_{B}}\;
\left(\frac{1}{p_{\pi}     \!\cdot\! v+ \delta \tilde{m}_{3} }-
\frac{1}{p_{\pi}     \!\cdot\! v^{\prime}- \delta \tilde{m}_{3} }
\right), \nonumber\\
f_{2 \, R}&=& \frac{M_{B}}{M_{D}}\;f_{ 1\, R} ,  \nonumber\\
f_{ 3\, R}&=& f_{ 4\, R}=0,  \nonumber\\
f_{ 5\, R}&=& \frac{\alpha_{1} \; \rho_{1}(\nu)}{2F_{0} M_{B} M_{D}}\;
\left(\frac{p_{\pi}     \!\cdot\! v}{p_{\pi}     \!\cdot\! v+ \delta
\tilde{m}_{1} }+\frac{p_{\pi}     \!\cdot\! v^{\prime}}
{p_{\pi}     \!\cdot\! v^{\prime}- \delta \tilde{m}_{1} }\right)
\nonumber\\
&+&  \frac{\alpha_{2} \; \rho_{2}(\nu)}{2F_{0} M_{B} M_{D}}\;
\left(\frac{p_{\pi}     \!\cdot\! v^{\prime}-\frac{1}{3} p_{\pi}
\!\cdot\! v \,(1+2\,\nu)}
{p_{\pi}     \!\cdot\! v+ \delta \tilde{m}_{2} }+
\frac{p_{\pi}     \!\cdot\! v-\frac{1}{3} p_{\pi}     \!\cdot\! v^{
\prime}\, (1+2\,\nu)}
{p_{\pi}     \!\cdot\! v^{\prime}- \delta \tilde{m}_{2} }\right)
\nonumber\\
&+&\frac{\alpha_{3} \;\xi^{(1)}(\nu)}{2 F_{0} M_{B} M_{D}}\;
\left( \frac{p_{\pi}     \!\cdot\! v}{p_{\pi}     \!\cdot\! v+ \delta
\tilde{m}_{3} }+
 \frac{p_{\pi}     \!\cdot\! v^{\prime}}{p_{\pi}     \!\cdot\! v^{
\prime}- \delta \tilde{m}_{3} }
\right),
\nonumber\\
f_{6 \, R}&=& \frac{ \alpha_{2}\;  \rho_{2}(\nu)\, (\nu-1)}
{6 F_{0}M_{B}}\; \left(\frac{1}{p_{\pi}     \!\cdot\! v+  \delta
\tilde{m}_{2} }-
\frac{1}{p_{\pi}     \!\cdot\! v^{\prime}- \delta \tilde{m}_{2} }
\right)\nonumber\\
& +& \frac{\alpha_{3}\; \xi^{(1)}(\nu)
}{2 F_{0} M_{B}}\; \left(\frac{1}{p_{\pi}     \!\cdot\! v+ \delta
\tilde{m}_{3} }-
\frac{1}{p_{\pi}     \!\cdot\! v^{\prime}- \delta \tilde{m}_{3} }
\right),\nonumber\\
k_{ R}&=&- \frac{\alpha_{1}\; \rho_{1}(\nu) \,(\nu-1)}{2F_{0}}\;
\left(\frac{p_{\pi}     \!\cdot\! v}{p_{\pi}     \!\cdot\! v+ \delta
\tilde{m}_{1} }+
\frac{p_{\pi}     \!\cdot\! v^{\prime}}{p_{\pi}     \!\cdot\! v^{
\prime}- \delta \tilde{m}_{1} }\right)\nonumber\\
&-& \frac{\alpha_{2} \; \rho_{2}(\nu)\,(\nu-1)}{3F_{0}}\;
\left(\frac{(p_{\pi}     \!\cdot\! v^{\prime}-\nu\; p_{\pi}     \!
\cdot\! v)}
{p_{\pi}     \!\cdot\! v+ \delta \tilde{m}_{2} }+
\frac{(p_{\pi}     \!\cdot\! v-\nu\; p_{\pi}     \!\cdot\! v^{
\prime})}
{p_{\pi}     \!\cdot\! v^{\prime}- \delta \tilde{m}_{2} }\right)
\nonumber\\
&+ &\frac{\alpha_{3}   \;  \xi^{(1)}(\nu)}{2F_{0}}\;
\left(\frac{(p_{\pi}     \!\cdot\! v^{\prime}-\nu\; p_{\pi}     \!
\cdot\! v)}
{p_{\pi}     \!\cdot\! v+ \delta \tilde{m}_{3} }+
\frac{(p_{\pi}     \!\cdot\! v -\nu\; p_{\pi}     \!\cdot\! v^{
\prime})}
{p_{\pi}     \!\cdot\! v^{\prime}- \delta \tilde{m}_{3} }\right),
\nonumber\\
g_{1 \, R}&=& 0, \nonumber\\
g_{2 \, R}&=&  -\frac{\alpha_{2}\; \rho_{2}(\nu)}{2 F_{0} \,M_{B}}
\frac{1}{p_{\pi}     \!\cdot\! v^{\prime}- \delta \tilde{m}_{2} },
\nonumber\\
g_{3 \, R}&=& - g_{2 \, R}, \nonumber\\
g_{ 4\, R}&=&\frac{\alpha_{2}\;  \rho_{2}(\nu)}{3 F_{0} \, M_{D}}\;
\left(\frac{2}{p_{\pi}     \!\cdot\! v+ \delta \tilde{m}_{2} }-(1+
\frac{1}{2}\,\nu)
\frac{1}{p_{\pi}     \!\cdot\! v^{\prime}- \delta \tilde{m}_{2} }
\right) \nonumber\\
& +&\frac{\alpha_{3} \; \xi^{(1)}(\nu)}{F_{0}}
\frac{1}{p_{\pi}     \!\cdot\! v^{\prime}- \delta \tilde{m}_{3} } ,
\nonumber\\
g_{5 \, R}&=&0 .
\end{eqnarray}

\end{document}